\documentclass{pasj01}
\draft 
\Received{$\langle$reception date$\rangle$}
\Accepted{$\langle$acception date$\rangle$}
\Published{$\langle$publication date$\rangle$}

\usepackage{lineno}
\usepackage{bm}
\def\icarus{\ref@jnl{Icarus}}           

\newcommand{\bfrac}[3]{\left(\frac{#1}{#2}\right)^{#3}}
\newcommand{\pdv}[2]{\frac{\partial #1}{\partial #2}}
\newcommand{\vel}{\bm{v}}
\newcommand{\unitt}{{\rm \bm{I}}}
\newcommand{\gpot}{\Phi}
\newcommand{\vistensor}{\bm{\Pi}}
\newcommand{\mstar}{M_{\ast}}
\newcommand{\mpl}{M_{\rm p}}

\newcommand{\apl}{a_{\rm p}}

\newcommand{\rpvec}{\bm{r}_{\rm p}}
\newcommand{\rvec}{\bm{r}}
\newcommand{\cs}{c_s}
\newcommand{\vk}{v_K}
\newcommand{\Omegak}{\Omega_K}
\newcommand{\hscale}{h}
\newcommand{\vradius}{v_r}
\newcommand{\vtheta}{v_{\theta}}
\newcommand{\vphi}{v_{\phi}}
\newcommand{\rhosurf}{\Sigma}
\newcommand{\iorb}{i_{\rm orb}}
\newcommand{\twco}{$^{12}$CO}
\newcommand{\thco}{$^{13}$CO}
\newcommand{\ceio}{C$^{18}$O}
\newenvironment{software}{\section*{Software}\fontsize{8}{11}\selectfont}{\par}
\newcommand{\rev}[1]{\color{black} #1 \color{black}}
\newcommand{\revs}[1]{\color{black} #1 \color{black}}

\begin{document}

\title{Kinematic signatures of a \rev{low-mass} planet with a \rev{moderately} inclined orbit in a protoplanetary disk}
\author{
   Kazuhiro D. Kanagawa\altaffilmark{1},
   Tomohiro Ono\altaffilmark{2},
   Munetake Momose\altaffilmark{1}
   }%
\altaffiltext{1}{College of Science, Ibaraki University, 2-1-1 Bunkyo, Mito, Ibaraki 310-8512, Japan}
\altaffiltext{2}{School of Natural Sciences, Institute for Advanced Study, Princeton, NJ 08544, USA}
\email{kanagawa.k@gmail.com}


\maketitle

\begin{abstract}
   A planet embedded in a protoplanetary disk produces a gap by disk--planet interaction. It also generates velocity perturbation of gas, which can also be observed as deviations from the Keplerian rotation in the channel map of molecular line emission, called kinematic planetary features. These observed signatures provide clues to determine the mass of the planet. We investigated the features induced by the planet with an inclined orbit through three-dimensional hydrodynamic simulations. We found that a smaller planet, with the inclination being $\sim 10^{\circ}$ -- $20^{\circ}$, can produce kinematic features as prominent as those induced by the massive coplanar planet.
   Despite the kinematic features being similar, the gap is shallower and narrower as compared with the case in which the kinematic features are formed by the coplanar planet. We also found that the kinematic features induced by the inclined planet were fainter for rarer CO isotopologues because the velocity perturbation is weaker at the  position closer to the midplane, which was different in the case with a coplanar massive planet. This dependence on the isotopologues is distinguished if the planet has the inclined orbit. We discussed two observed kinematic features in the disk of HD~163296. We concluded that the \rev{kink observed at 220~au} can be induced by the inclined planet, while the \rev{kink at 67~au} is consistent to that induced by the coplanar planet.
\end{abstract}

\section{Introduction} \label{sec:introduction}
A planet is formed in a protoplanetary disk and gravitationally interacts with the surrounding gas via spiral waves \citep{Lin_Papaloizou1979,Goldreich_Tremaine1980}.
When a planet is sufficiently large, a gas depleted region is formed along with the planet orbit, called a planetary gap (e.g.,\cite{Lin_Papaloizou1986a}, \cite{Kley1999}, \cite{Crida_Morbidelli_Masset2006}, \cite{Duffell_MacFadyen2013}, \cite{Fung_Shi_Chiang2014}, \cite{Kanagawa2015a}).
Recent observations with Atacama Large Millimeter/submillimeter Array (ALMA) have revealed a number of protoplanetary disks with gap structures (e.g., \cite{ALMA_HLTau2015}, \cite{Tsukagoshi2016}, \cite{Long2018}, \cite{DSHARP1}, \cite{VanderMarel2019}, \rev{\cite{Cieza2019}).}
The planetary gap is one of the convincing origins to explain these observed gap structures, although other processes (e.g., the dust evolution in the vicinity of the snow line \citep{Zhang_Blake_Bergin2015,Okuzumi_Momose_Sirono_Kobayashi_Tanaka2016}, secular gravitational instabilities \citep{Takahashi_Inutsuka2014,Takahashi_Inutsuka2016a}, disk-wind \citep{Takahashi_Muto2018}) are proposed.
Moreover, deviations from the Keplerian rotation have been detected in the channel maps of \twco~ emissions in some protoplanetary disks (e.g., \cite{Pinte_Price2018}, \cite{Teague_Bae_Bergin_Birnstiel_Foreman2018}, \rev{\cite{Casassus_Perez2019}, \cite{Teague2019}, \cite{Pinte2019}}, \cite{Pinte2020}, \cite{Bae_AS209_2022}, \rev{\cite{Izquierdo2022, Calcino2022}, and see also \cite{Pinte_PPVII}).}
These anomalies from the Keplerian rotation are thought to be originated from the velocity perturbation induced by the embedded planet \rev{\citep{Perez2015,Pinte_Price2018} and density perturbation (spiral and gap) in addition to the velocity perturbations \citep{Perez_Casassus_Benitez_Llmabay2018,Rabago_Zhu2021}.}

Particularly, \cite{Pinte2020} found the kinematic planetary features (or called velocity kinks) that can be generated from the planet within the observed gap, in nine protoplanetary disks in the DSHARP sample \citep{DSHARP1}, which strongly suggests that the observed gap is induced by the planet.
However, they also highlighted that the planet mass suggested by the kinematic planetary feature was 4--10 times larger than that estimated by the gap shape.
To address this discrepancy of the masses predicted by the gap and kinematic planetary feature, we consider a planet with an inclined orbit in this paper.

Exoplanets with misalignment of the angular momentum vector with the stellar spin axis have been discovered \citep{Winn_Fabrycky2015}.
This misalignment can indicate that the planet has an inclined orbit against the protoplanetary disk due to, for instance, the planet-- scattering in the protoplanetary disk (e.g.,\cite{Chatterjee2008}).
\rev{
The disk--planet interaction for the low-mass planet with low-inclination has been investigated by using a linear theory \citep{Tanaka_Ward2004} and hydrodynamic simulations (e.g., \cite{Arzamasskiy2018}).
In the case with the massive planet, it has been investigated by hydrodynamic simulations (e.g., \cite{Cresswell_Dirksen_Kley_Nelson2007}, \cite{Marzari_Nelson2009}, \cite{Bitsch_Kley2011}, \cite{Xiang_Gruess_Papaloizou2013}, \cite{Kloster_Flock2019}, \cite{Zhu2019}).
}
With the orbital inclination, a larger mass of a planet is required to form a gap \citep{Bitsch_Crida_Libert_Lega2013}.
Moreover, the planet could disturb gas at a high altitude compared with the case with the coplanar planet, which may appear in the structure of the channel map as the kinematic planetary features.

In this paper, we investigate the kinematic planetary features formed by the \rev{low-mass} planet with \rev{moderate orbital inclination} using three-dimensional (3D) hydrodynamic simulations and radiative transfer calculations.
We present our model and numerical method in Section~\ref{sec:method}.
In Section~\ref{sec:results}, we present our results and discuss the implications of our results to observations and possible scenarios to maintain the orbital inclination in the disk phase in Section~\ref{sec:discussion}.
Finally, the summary of our results is provided in Section~\ref{sec:summary}.

\section{Our model and method} \label{sec:method}
\subsection{Basic equations}
We considered 3D hydrodynamics of a protoplanetary disk with an inclined-orbital planet, assuming a non-self-gravitational and locally isothermal disk.
Hence, the basic equations are expressed as
\begin{eqnarray}
   &\pdv{\rho}{t} + \nabla \cdot \left( \rho \vel \right) = 0 \label{eq:cont}\\
   &\pdv{\rho \vel}{t} + \nabla \cdot \left(\rho \vel + P \unitt\right) = -\rho \nabla \gpot + \nabla \cdot \vistensor, \label{eq:eom}
\end{eqnarray}
where $\rho$ and $\vel$ are the gas density and velocity, respectively. 
We used the spherical coordinates, $r$ is the spherical radius, $\theta$ and $\phi$ are the polar and azimuth angles, respectively, and hence $\rvec$ = (r,$\theta$,$\phi$) and $\vel$=($\vradius$,$\vtheta$,$\vphi$).
The origin of the coordinate is set to be the position of the central star.
For simplicity, note that we also used the cylindrical coordinates (R, $\phi$, z), where $R=r\sin (\theta)$ and $z=r\cos(\theta)$.
The unit tensor is denoted by $\unitt$.
Since we adopted the simple locally isothermal equation of state, the gas pressure $P$ is expressed as 
\begin{eqnarray}
   P &= \cs^2 \rho, \label{eq:locally_isothermal_eos}
\end{eqnarray}
where $\cs$ is the sonic speed, which is given by a function of $R$. 
The gravitational potential $\gpot$ is expressed as
\begin{eqnarray}
   \gpot &= -\frac{G\mstar}{\left| \rvec \right|} - \frac{G\mpl}{\sqrt{\left(\left| \rvec - \rpvec \right|\right)^2 + \epsilon^2}} + \frac{G\mpl}{\left| \rpvec \right|^3}\rvec \cdot \rpvec,
   \label{eq:gpot}
\end{eqnarray}
where $G$ is the gravitational constant and $\mstar$ and $\mpl$ are the masses of the central star and planet, respectively.
The plant is located at $\rvec=\rpvec$.
In the following, subscript p denotes the quantities of the planet.
The smoothing parameter of planet gravity $\epsilon$ is set as $0.1 r_H$, where $r_H$ is the hill radius of the planet defined by $\apl \left[\mpl/(3\mstar)\right]^{1/3}$ in which $\apl$ is the semimajor axis of the planet.
The viscous stress tensor $\vistensor$ is 
\begin{eqnarray}
   \vistensor &= \rho \nu \left[ \nabla \vel + \left(\nabla \vel\right)^{\dag} -\frac{2}{3} \unitt \nabla \cdot \vel \right], \label{eq:vistensor}
\end{eqnarray}
where $\nu$ is the kinetic viscosity.
We adopted the $\alpha$-viscosity prescription (\cite{Shakura_Sunyaev1973}). Hence,
\begin{eqnarray}
   \nu &= \alpha \cs \hscale ,
   \label{eq:alphavis}
\end{eqnarray}
where $\hscale$ is the scale height ($=\cs/\Omegak$) and $\Omegak$ is the Keplerian angular velocity defined by $\sqrt{G\mstar/R^3}$, and \revs{we use $\alpha=10^{-3}$ in this paper}.
For convenience, we also defined $\vk=R\Omegak$ as the Keplerian rotation velocity.

\subsection{Initial and boundary conditions}
We assumed a power-law distribution of the initial surface density $\rhosurf$, which is expressed as 
\begin{eqnarray}
   \rhosurf &= \rhosurf_0 \bfrac{R}{R_0}{p_s}, \label{eq:rhosurf_init}
\end{eqnarray}
where $R_0$ is the unit of the radius. We adopted $p_s=-1/2$ and $\rhosurf_0=1$ in the following.
Note that the choice of $\rhosurf_0$ does not affect the result of the hydrodynamic simulation, except for the absolute value of the density, since we do not consider the self-gravity of the gas disk.
We also used $\mstar$ as the unit of the mass and $\Omegak^{-1}(R_0)$ as the unit of time; hence, $\rhosurf$ and $\rho$ are normalized by $\mstar/R_0^2$ and $\mstar/R_0^3$, respectively.
We also assumed the power-law distribution of the sound speed and vertically isothermal distribution, which is expressed as
\begin{eqnarray}
   \cs &= \cs{}_0 \bfrac{R}{R_0}{q}, \label{eq:csdist}
\end{eqnarray}
where we set $q=-1/2$ in this paper.
The aspect ratio of the disk, $H$, is written by $\cs/\vk$, and it is constant as $H_0=\cs{}_0/\vk(R_0)$ throughout the computational domain. 
\rev{We investigated the cases with $H_0=0.07$ and $H_0=0.1$, referring to the recent observations of the outer region of the planetary disks (e.g., \cite{Rich2021}, \cite{MAPS1}).}
Assuming hydrostatic equilibrium, we adopted the Gaussian profile in vertical direction, which is expressed as 
\begin{eqnarray}
   \rho &= \frac{\rhosurf_0}{\sqrt{2\pi} \hscale{}_0} \bfrac{R}{R_0}{p}\exp\left[ \frac{G\mstar}{\cs^2} \left( \frac{1}{r} - \frac{1}{R} \right) \right],
   \label{eq:init_rhodist}
\end{eqnarray}
where $\hscale{}_0$ is the scale height at $R_0$ and $p=p_s-(2q+3)/2$ is equal as $-1.5$ in this paper.

The initial azimuthal velocity is given by hydrodynamic equilibrium in $R$ direction; that is,
\begin{eqnarray}
   \Omega &= \Omegak \left[ \left( 1+q \right) + \left( p+q \right)H^2 - q\frac{R}{r} \right].
    \label{eq:init_omega}
\end{eqnarray}
The viscous drift velocity is given by 
\begin{eqnarray}
   v_{\rm vis} &= -\frac{3\nu}{R} \left[p+ \frac{4}{3}q +\frac{10q+9}{3}\frac{G\mstar}{\cs^2} \left(\frac{1}{r} - \frac{1}{R} \right) \right].
   \label{eq:vis_velocity}
\end{eqnarray}
We assumed vertical hydrostatic equilibrium $v_z=0$.
Therefore, $\vradius=v_{\rm vis} \sin(\theta)$ and $\vtheta=v_{\rm vis} \cos (\theta)$ in the initial profile.

At the inner and outer boundaries in the $r$ direction, we fixed the physical quantities on the initial values.
At the upper and lower boundaries in the $\theta$ direction, we adopted a vertically outgoing boundary for $\vradius$ and $\vtheta$.
That is, the velocities are copied just inside of the boundary if the gas flows outside of the computational domain ($v_z > 0$ and $z>0$, or $v_z < 0$ and $z<0$); otherwise, $\vradius$ and $\vtheta$ are set to zero.
The azimuthal velocity is copied from that of the cell just next to the boundary.
Note that we applied no wave-damping zone for simplicity.

\subsection{Numerical methods}
\subsubsection{Hydrodynamic simulations}
We used Athena++\footnote{https://www.athena-astro.app} \citep{Athena++} to solve basic equations (\ref{eq:cont}) and (\ref{eq:eom}).
Athena++ is a grid-based magnetohydrodynamic code using a higher-order Godunov scheme with the orbital advection scheme (\cite{Masset2000}, see Ono et al. in preparation) and static/adaptive mesh refinement.

The computational domain runs from $r=0.4R_0$ to $r=3.0R_0$ in $r$-direction and covers the region of $- 5.6 H_0 < \theta-\pi/2 < 5.6 H_0$ and $-\pi < \phi < \pi$.
We logarithmically divided the computational domain into 192 meshes in the $r$-direction and equally divided it by 80 meshes in the $\theta$-direction and 640 meshes in the $\phi$ direction.
In addition, we used one-level static mesh refinement in the range of $0.6 R_0 < r < 1.4 R_0$, and the root grid was divided into two in the refinement region.
The grid size is approximately $5\times 10^{-3}$ in the region where the refinement is adopted.
When $H_0=0.07$, which is our fiducial value, the scale height is divided into 14 meshes in the refinement region, while it is divided in 7 meshes in the other region. 

We adopted the second-order piecewise linear method for the spatial reconstruction method (it is the default of ATHENA++) and third-order accurate Strong Stability Preserving (SSP) variant method for time-integration with the CFL coefficient being 0.6.
Note that we confirmed that the result does not change if we chose the second-order van Leer scheme (the default of ATHENA++) with CFL coefficient being 0.3.

\subsubsection{Planet}
We considered a planet with an orbital inclination.
For simplicity, we ignored a torque from the surrounding gas and planetary migration.
We only considered the gravity of the central star when calculating the planetary orbit.
The initial location of the planet was set by $\rpvec=(R_0, \pi/2-\sin\left(\iorb\right) \sin \left(\pi/4\right), 0)$, and the initial velocity was set by $\bm{v}_p=(0, \sin\left(\iorb\right)\sin\left(\pi/4\right),\vk(R_0))$, where $\iorb$ is the orbital inclination of the planet\footnote{\rev{The initial condition is a bit different from the steady state condition, and hence the planet move slight from the initial radius. However, the planet reaches steady state after just after a few orbits. Hence, it could not affect our result significantly.}}.
The time integration for the orbital calculation was used by the second-order leapfrog method to stably calculate the planetary orbit long time.

We increased the planet mass gradually in the beginning using
\begin{eqnarray}
   \mpl &= \mpl{}_{\rm ,fin} \sin^2 \left( \frac{\pi}{2} \frac{t}{\tau_{m}}\right),
   \label{eq:mpl_ramp}
\end{eqnarray}
until $t=\tau_m$, where \rev{$\tau_m$ is the timescale of the mass growth and} we adopted $\tau_m = 2\pi/\Omegak(R_0)$.
After $t=\tau_m$, we fixed the planetary mass on $\mpl{}_{\rm , fin}$.
In the following, we referred $\mpl{}_{\rm ,fin}$ as $\mpl$.

\subsubsection{Radiative transfer simulations}
We used the density and velocity profiles given by the 3D hydrodynamic simulations described above as an input model, but is slightly modified (see below), to RADMC3D\footnote{https://www.ita.uni-heidelberg.de/~dullemond/software/radmc-3d} \citep{RADMC3D} that is a publicly available code for radiative transfer calculations for dust and molecules.
In the following, we used the results of hydrodynamic simulations at $t=100$~orbits, corresponding to $0.1$~Myr when $R_0$=100~au.

First, we calculated the gas temperature by carrying out thermal Monte Carlo simulation by assuming that the gas temperature is equal to the dust temperature.
Then, we carried out molecular line transfer simulations to produce synthetic channel maps using the temperature produced by the thermal Monte Carlo.
Note that the temperature adopted in the radiative transfer simulations is not consistent to that adopted in the 3D hydrodynamic simulations with the locally isothermal equation of state.
This temperature inconsistency may not affect our main result described in Section~\ref{sec:results} (see more discussion in Appendix~\ref{sec:channel_maps}).

We assumed a T-Tauri \rev{star} that is similar to a protostar as a central star with $\mstar=1M_{\odot}$, and its radius is $2.5 R_{\odot}$.
The effective temperature of the star was set to be 4500~K; hence, its bolometric luminosity was approximately $1.7L_{\odot}$.
The star is located in the origin of the coordinate the same as in the hydrodynamic simulation.

We assumed that the dust grains are well-mixed to gas; hence, for simplicity, the scale height of the dust grains is the same as that of the gas.
The size distribution of the dust grains is assumed to be $s^{-3.5}$, where $s$ is the size of the dust grain.
The minimum and maximum sizes of the dust grain are $0.1$~$\mu$m and $1$~mm, respectively.
We used the same compositions of the dust grains to that adopted in DSHARP \citep{DSHARP_DUST}, and the absorption and scattering coefficients are averaged by the size distribution.
We assumed that the gas-to-dust mass ratio is 100, and it is constant in the entire domain.

In this paper, we considered the rotation transition in $J=2-1$ of \twco~ and its isotopologues \thco~ and \ceio.
The fractional abundance of \twco~ relative to H$_2$ was set to $9.83\times 10^{-5}$, and the abundances of \thco~ and \ceio~ were $1.49 \times 10^{-6}$ and $2.26 \times 10^{-7}$, respectively.
The line transfer was done in local thermal equilibrium using molecular data from the LAMBDA database\footnote{https://home.strw.leidenuniv.nl/~moldata/} \citep{Schoier2005,vanderTak2020}.

We adopted $10^7$ photons for the thermal Monte Carlo simulations and  $3\times 10^{6}$ photons for imaging of dust continuum and molecular line emissions, respectively.
The density and velocity structures in the radiative transfer were fed from the 3D hydrodynamic simulation with $R_0=100$~au. However, the profile has a relatively large cavity because of the inner boundary of the computational domain (its size was 40~au when $R_0=100$~au).
This unintentional cavity can slightly affect the temperature and channel map.
With this, we inserted a Keplerian rotation disk profile to extend the initial profile from the inner boundary of the hydrodynamic simulation to $10$~au (=$0.1R_0$).
\revs{We also should emphasis that the structure close to the inner boundary of the simulation, that is $\sim 40$~au, can be affected by the boundary condition of the simulation.
To address the inner structure, e.g., inner secondary gap (e.g., \cite{Dong_Li_Chiang_Li2017},\cite{Bae_Zhu_Hartmann2017}), a simulation with the inner boundary at smaller radii is required. 
These structures are out of the scope of this paper.}

We assumed that the distance from the star was 140~pc and the inclination and position angles were $45^{\circ}$ and $135^{\circ}$, respectively. 
For the initial disk mass, we assumed to be $0.1M_{\odot}$.
The planet is located at ($x_{\rm p}, y_{\rm p}$) = (-0.06~arcsec, 0.56~arcsec)  or (-8.4~au,78.4~au )in the images.

\section{Results} \label{sec:results}
\subsection{CO emission layers} \label{subsec:em_layers}
Before showing the results with the planet, we presented a height of the emission layer of CO isotopologues in the disk with no planet because it reveals the region that mainly contributes in the determination of the structure of the channel map.
\begin{figure}
   \begin{center}
      \resizebox{0.49\textwidth}{!}{\includegraphics{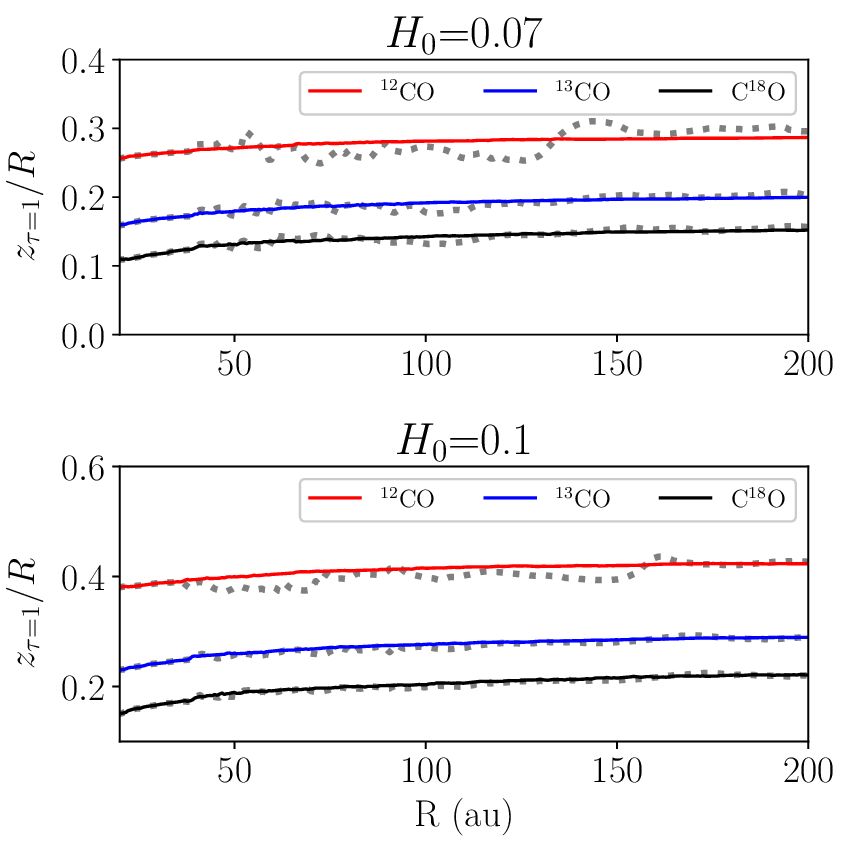}}
   \end{center}
   \caption{
      Locations of $\tau=1$ for $J=$2-1 emissions of \twco, \thco, and \ceio for the unperturbed disk (solid lines) with $H_0=0.07$ (top) and $H_0=0.1$ (bottom). The dotted lines represent the case with the planet (Run 3 in Table~\ref{tab:params}).
   \label{fig:tausurf1}}
\end{figure}
Figure~\ref{fig:tausurf1} shows the emission layers of \twco, \thco, and \ceio, for the cases with $H_0=0.07$ and $0.1$.
We defined the emission layer as a location of $\tau=1$ in the face-on disk, which is calculated by RADMC3D, where $\tau$ is the optical depth.
For reference, we plotted the emission layers in the case with the inclined planet (Run~3 in Table~\ref{tab:params}) in the figure.
There is no significant difference between the cases with/without planet because the planet only forms a shallow gap (see next subsection).
The emission layers shift down close to the midplane when the planet forms a deep gap (Appendix~\ref{sec:coplanar_case}).

The ratio of the height of the emission layer $z_{\tau=1}$ to $R$ is approximately constant in the simulation domain (Figure~\ref{fig:tausurf1}).
In the case with $H_0=0.07$ and $R_0=100$~au, $z_{\tau=1}/R_0 = 0.29 \simeq 4H_0$ for \twco, $z_{\tau=1}/R_0 = 0.21 \simeq 3H_0$ for \thco, and $z_{\tau=1}/R_0=0.11 \simeq 1.6H_0$ for \ceio.
When $H_0=0.1$, the emission layers for \twco~and \thco~are $4H_0$ and $3H_0$, respectively, following the similar scaling as in the case with $H_0=0.07$.
However, for \ceio, $z_{\tau=1}/R_0=2.3H_0$ when $H_0=0.1$, and it is slightly larger than that in the case of $H_0=0.07$.
In all cases, the emission layers are always contained within the computational domain of hydrodynamic and radiative transfer simulations.

\subsection{Kinematic features in channel maps with varying planetary orbital inclination}
We simulated the disk structure with a planet with planet mass $\mpl$ and orbital inclination $\iorb$ (Table~\ref{tab:params}).
We chose $H_0=0.07$ as our fiducial value (Runs 1--6) and demonstrated the cases with $H_0=0.1$ for $\mpl/\mstar=5\times 10^{-4}$ (Runs 7-10) to see a dependence on the aspect ratio.
For reference, we also carried out the simulation with a massive planet ($\mpl/\mstar=2\times 10^{-3}$) with the coplanar orbit (Run~11).
We used $\alpha=10^{-3}$ in all the cases.
\begin{table}
   \tbl{Parameters}{%
   \begin{tabular}{clcccc}
   \hline\noalign{\vskip3pt}
   Run& $\mpl/\mstar$ & $H_0$ & $\alpha$ &$\iorb$  \\ 
      &               &       &          &(degree) \\  [2pt]
   \hline\noalign{\vskip3pt}
   1& $5\times 10^{-4}$ ($0.5M_J$)$^{*}$& $0.07$ & $10^{-3}$ & $0.0$  \\
   2& $5\times 10^{-4}$           & $0.07$ & $10^{-3}$ & $7.1$ \\
   3& $5\times 10^{-4}$           & $0.07$ & $10^{-3}$ & $14.1$ \\
   4& $5\times 10^{-4}$           & $0.07$ & $10^{-3}$ & $21.1$ \\
   \hline\noalign{\vskip3pt}
   5& $3\times 10^{-4}$  ($0.3M_J$)& $0.07$ & $10^{-3}$ & $14.1$ \\
   6& $1\times 10^{-4}$  ($0.1M_J$)& $0.07$ & $10^{-3}$ & $14.1$ \\
   \hline\noalign{\vskip3pt}
   7& $5\times 10^{-4}$  ($0.5M_J$)& $0.10$ & $10^{-3}$ & $0.0$  \\ 
   8& $5\times 10^{-4}$            & $0.10$ & $10^{-3}$ & $14.1$ \\ 
   9& $5\times 10^{-4}$            & $0.10$ & $10^{-3}$ & $21.1$ \\ 
   10& $5\times 10^{-4}$           & $0.10$ & $10^{-3}$ & $29.4$ \\ 
   \hline\noalign{\vskip3pt}
   11& $2\times 10^{-3}$  ($2.0M_J$)& $0.07$ & $10^{-3}$ & $0.0$ \\ [2pt]
   \hline\noalign{\vskip3pt}
   \end{tabular}} \label{tab:params}
   \rev{$^{*}$ Mass of the planet in Jupiter mass when $M_{\ast} = 1 M_{\odot}$. }
\end{table}

\begin{figure*}
   \begin{center}
      \resizebox{0.98\textwidth}{!}{\includegraphics{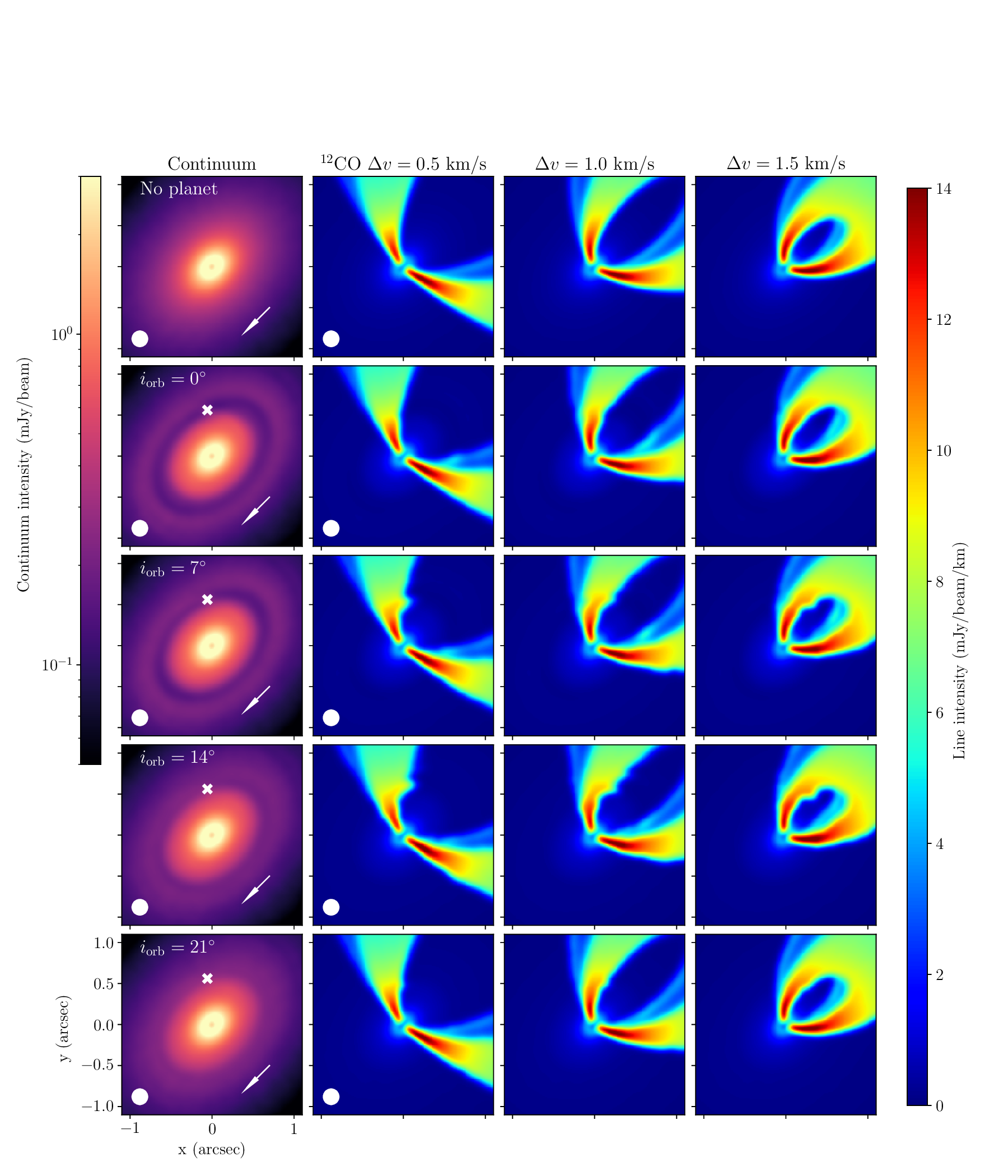}}
   \end{center}
   \caption{
      The dust continuum images at 230~GHz (left) and the channel maps of \twco~J=2-1 emission for various $\iorb$ in the cases with $\mpl/\mstar=5\times 10^{-4}$ (Runs 1--4 in Table~\ref{tab:params}). 
      The uppermost panels are the case of the simple Keplerian disk with no planet, and the subsequent lines correspond to the cases with the planet with $\iorb=0^{\circ}$, $7.1^{\circ}$, $14.1^{\circ}$, and $21.1^{\circ}$.
      The white cross in the dust continuum image represents the location of the planet.
      \rev{The disk rotation is clockwise and the white arrow in the lower right corner of the continuum images indicate the direction of the rotation. In the following figures, the direction of the disk rotation is the same.}
   \label{fig:cont_co_q5e-4_ivar}
   }
\end{figure*}
Figure~\ref{fig:cont_co_q5e-4_ivar} shows the dust continuum images at 230~GHz and the channel maps of \twco~J=2-1 emission when $\mpl/\mstar=5\times 10^{-4}$ with various $\iorb$.
The meridian angle of the planet varies with time when the orbital inclination is nonzero.
In the figure, the planet is located at the midplane ($z_{\rm p}$=0).
For reference, we show the case with no planet, demonstrating the pattern of the purely Keplerian rotation disk (the effect of the pressure gradient is included, as expressed in Equation~\ref{eq:init_omega}), in which there is no signature due to, for instance, vertical shear instability as discussed by \cite{Barraza-Alfaro_Flock_Marino_Ferez2021} because of the relatively large viscosity we adopted \rev{and our coarse resolution in \revs{vertical} direction \citep{Flores-Rivera2020}}. 
When $\iorb=0^{\circ}$, the relatively deep gap is observed in the dust continuum, and the channel map is similar to that for the Keplerian rotation disk with no planet, although the small deviations can be observed in some points, particularly in $\Delta v=1$~km/s map.
\rev{The} gap becomes shallower and narrower as $\iorb$ increases.
Moreover, different from the case with $\iorb=0^{\circ}$, significant deviations from the Keplerian rotation are observed around the planet in the channel map in the cases of $\iorb=7^{\circ}$ and $\iorb=14^{\circ}$.
When the inclination is large and $\iorb=21^{\circ}$, the deviation in the channel map becomes as weak as that in the case of $\iorb=0^{\circ}$, whereas almost no gap can be observed in the dust continuum.

\begin{figure*}
   \begin{center}
      \resizebox{0.98\textwidth}{!}{\includegraphics{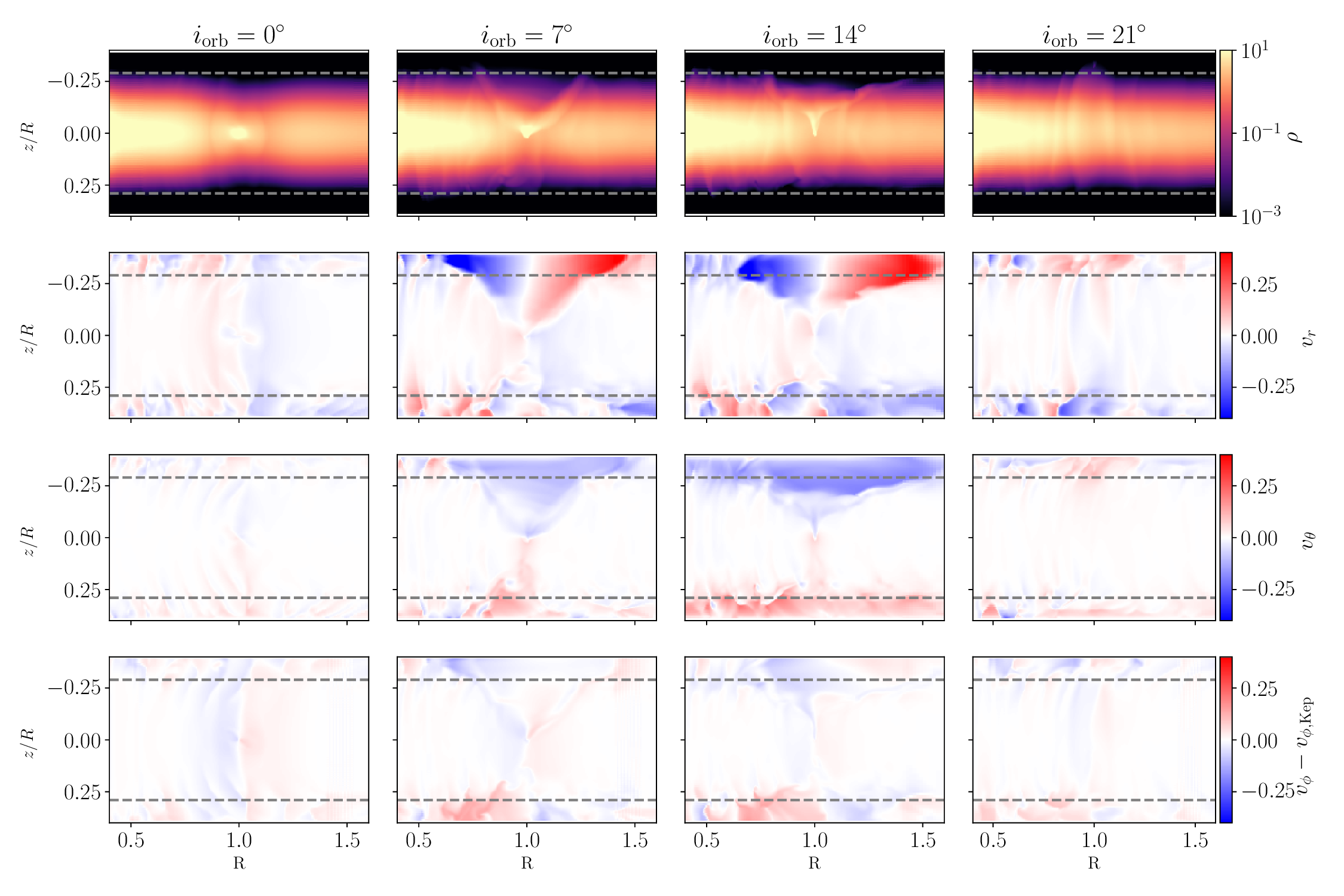}}
   \end{center}
   \caption{
      \rev{Density ($\rho$) and perturbed} velocity ($\vradius$, $\vtheta$, $\vphi - \vphi{}_{,\rm Kep}$) distributions that are averaged in the range of $\phi = \phi_{\rm p} \pm R_H/R_0$, which is the same cases shown in Figure~\ref{fig:cont_co_q5e-4_ivar}, except for the Keplerian case with no planet.
      The horizontal dashed lines indicate the emission layer of \twco~(see Figure~\ref{fig:tausurf1}).
      Note that the front layer of Figure~\ref{fig:tausurf1} corresponds the layer of $z<0$ since we adopted $i_{\rm disk}=135^{\circ}$. 
   \label{fig:velocity_dist_q5e-4}
   }
 \end{figure*}
 In Figure~\ref{fig:velocity_dist_q5e-4}, we show the velocities that are averaged in the vicinity of the planet location (in the range of $\phi=\phi_{\rm p} \pm R_H/R_0$) in the cases shown in Figure~\ref{fig:cont_co_q5e-4_ivar}.
 When $\iorb=0^{\circ}$, the gas velocity around the emission layer is small because it is distant from the planet.
 Similarly, the velocity around the emission layer is not significant when $\iorb=21^{\circ}$.
 Meanwhile, with a moderate inclination, such as $\iorb=7^{\circ}$ and $14^{\circ}$, the velocities at a high altitude are significantly disturbed by the planet.
 Particularly, the perturbation of $\vradius$ is prominent, and its amplitude is approximately 30~\% of the Keplerian rotation velocity.
 The disturbance of $\vradius$ causes the deviation from the Keplerian rotation appeared in the channel map shown in Figure~\ref{fig:cont_co_q5e-4_ivar} (see also Appendix~\ref{sec:vel_and_kps}).
 Note that the velocity distribution and relevant kinematic features induced by the inclined planet vary in the orbital period of the planet, which is further discussed in Section~\ref{subsec:location_dependence}

\rev{
\begin{figure*}
   \begin{center}
      \resizebox{0.98\textwidth}{!}{\includegraphics{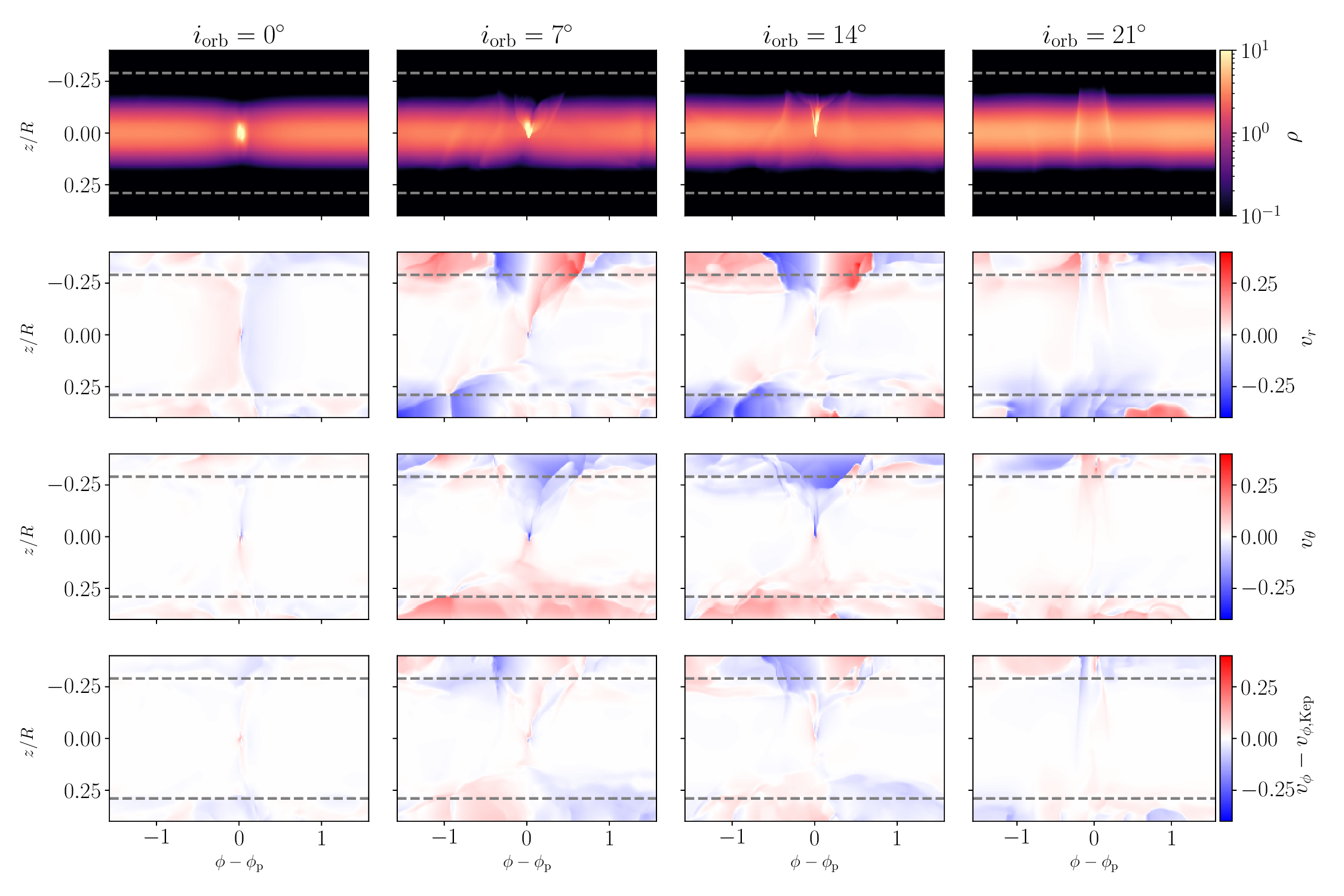}}
   \end{center}
   \caption{
   \rev{
   Azimuthal distribution of density and perturbed velocities around $R=R_p$ (averaged in the range of $R_p \pm R_H$), in the same case as shown in Figure~\ref{fig:velocity_dist_q5e-4}.
   }
   \label{fig:velocity_yzdist_q5e-4}
   }
 \end{figure*}
 Figure~\ref{fig:velocity_yzdist_q5e-4} shows the azimuthal distributions of density and perturbed velocities around $R=R_p$, in the same case of Figure~\ref{fig:velocity_dist_q5e-4}.
 As shown in Figures~\ref{fig:velocity_dist_q5e-4} and \ref{fig:velocity_yzdist_q5e-4}, one finds corn-like shape of $\rho$ and $v_r$, $\vphi$, which corresponds to the mach corn induced by the planet.
 Hence, we conclude that the kinematic features shown in Figure~\ref{fig:cont_co_q5e-4_ivar} are reflected from the planetary mach corn.  
 The corn is thinner with larger inclination angle and higher vertical velocity of the planet, and it is almost vanished in the case of $i_{\rm orb}=21^{^{\circ}}$.
}

\subsection{Mass dependence and variations of kinematic features with CO isotopes} \label{subsec:mass_dependence}
\begin{figure*}
   \begin{center}
      \resizebox{0.98\textwidth}{!}{\includegraphics{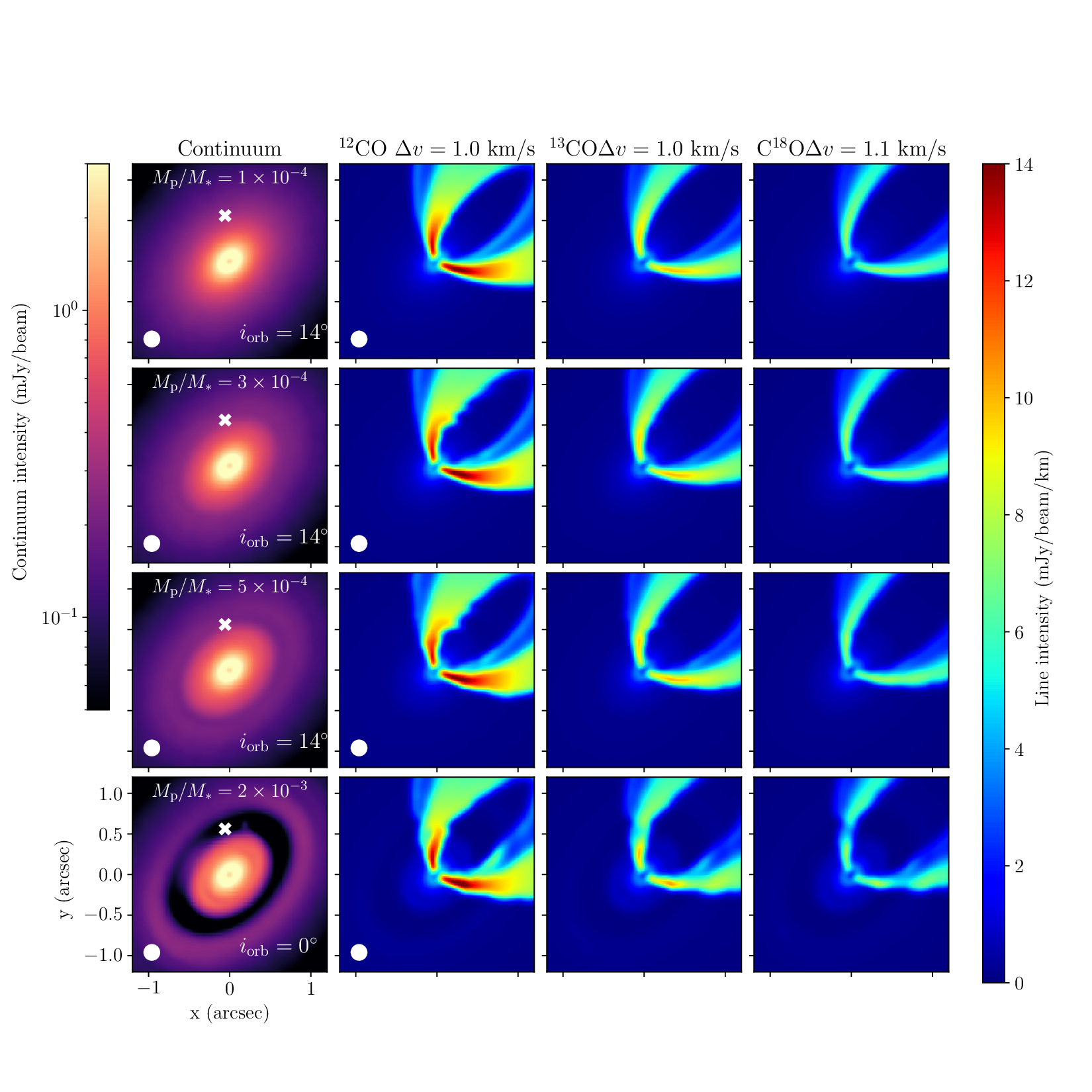}}
   \end{center}
   \caption{
      Dust continuum at 230~GHz (left) and the channel maps for $\Delta v=1$~km/s of \twco, \thco, and \ceio~ for various planet masses.
      Except for the panels in the bottom line, the orbital inclination is $14^{\circ}$.
      For reference, the bottom line shows the case with the coplanar massive planet ($\mpl/\mstar=2\times 10^{-3}$ and $\iorb=0^{\circ}$).
      \revs{In all the panels, the planet locates at the midplane.}
   \label{fig:cont_coisotops_qvar}
   }
 \end{figure*}
Figure~\ref{fig:cont_coisotops_qvar} shows the parts of the channel maps for the CO isotopologues (\twco, \thco, and \ceio) with various planet mass with $\iorb=14^{\circ}$ (Runs 4, 5, and 6), except for the panels in the bottom line.
In the bottom line of the figure, we show the channel maps in the case with the massive planet in the coplanar orbit ($\mpl/\mstar=2\times 10^{-3}$ and $\iorb=0^{\circ}$, Run 11), for reference.
\revs{Note that the planet locates at the midplane in all cases shown in the figure, as the same as Figure~\ref{fig:cont_co_q5e-4_ivar}}

In the case of $\mpl/\mstar=10^{-4}$, almost no gap is formed in the dust continuum, and no feature of the planet is observed in all the channel maps of the isotopologues.
As the mass increases, the dust continuum gap is gradually prominent, and some features associated to the planet can be observed in the \twco~ channel map.
However, there is no significant feature in the channel maps of other isotopologues, including \thco~ and \ceio.
The emission layers of \thco~ and \ceio~ are located closer to the midplane as compared to that of \twco, as presented in Section~\ref{subsec:em_layers}.
Considering the velocity perturbation shown in Figure~\ref{fig:velocity_dist_q5e-4}, the gas disturbance is weaker as it is close to the midplane.
No significant deviation appears in the channel maps of those molecules since the motion of gas near the emission layers of \thco~ and \ceio~ is approximately Keplerian motion. 

In contrast, in the case with the coplanar massive planet shown in the bottom line of Figure~\ref{fig:cont_coisotops_qvar}, the deep and wide gap is observed in the dust continuum.
In the channel map of \twco, the significant anomaly from the Keplerian rotation can be found.
In the channel map of \thco, the emission in the vicinity of the planet becomes a bit fainter because of gap formation, and the neck-like structure can be observed.
In the \ceio~map, the emission around the planet is much fainter than other parts, and then the gap structure is formed.
The neck and gap structures in the channel map also appear in the near and far sides that the planet is located.
These structures are produced by the density perturbation, rather than velocity perturbation, as discussed in Appendix~\ref{sec:vel_and_kps}.
The appearance of the substructures in the rare isotopologues is one of distinguishing features between the coplanar massive planet and the small planet with an inclined orbit.

\revs{Note that one may see the inner secondary gap in the case of $\mpl/\mstar  = 2\times 10^{-3}$ (the bottom line of Figure~\ref{fig:cont_coisotops_qvar}). However, it locates close to the inner boundary of the 3D simulation. 
Hence it could be just the artifact of the simulation.
A simulation with the inner boundary at smaller radii is needed to investigate this issue.
}

\subsection{Case with a larger disk scale height ($H_0=0.1$)} \label{subsec:higher_scaleheigh_case}
\begin{figure*}
   \begin{center}
      \resizebox{0.98\textwidth}{!}{\includegraphics{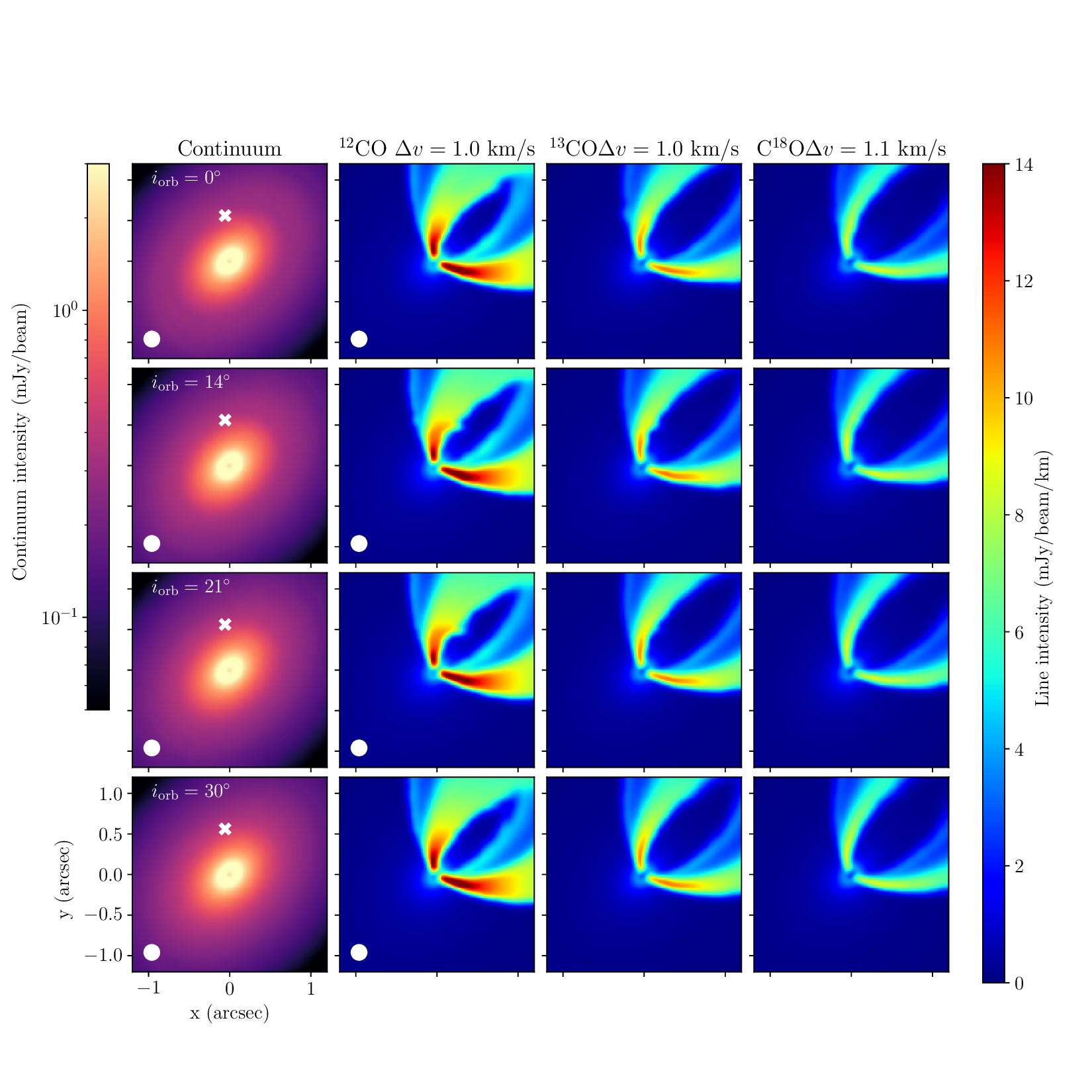}}
   \end{center}
   \caption{
      The same as in Figure~\ref{fig:cont_coisotops_qvar}, but for the case with $\mpl/\mstar=5\times 10^{-4}$, $H_0=0.1$ and various $\iorb$ (Run 7--10).
   \label{fig:cont_coisotops_h01_ivar}
   }
 \end{figure*}
 We carried out simulations with higher $H_0$ (=0.1) and various orbital inclination (Runs 7--10) to determine the dependence on the disk aspect ratio.
 The results are shown in Figure~\ref{fig:cont_coisotops_h01_ivar}.
 There is no visible gap in the dust continuum in all the cases shown in the figure.
 However, in the channel maps, a significant anomaly from the Keplerian disk is observed when $\iorb=14^{\circ}$ and $21^{\circ}$.
 In the case with $H_0=0.1$, the emission layer lies at a higher altitude compared to the case of $H_0=0.07$ (Figure~\ref{fig:tausurf1}).
 Considering this, the deviation from the Keplerian rotation in the channel map of \twco~ can be observed with a larger orbital inclination.
 Similar to the case with $H_0=0.07$, the deviation becomes weaker and vanishes for rarer isotopologues.

 Considering the results with $H_0=0.07$ (Figure~\ref{fig:cont_coisotops_qvar}) and $H_0=0.1$ (Figure~\ref{fig:cont_coisotops_h01_ivar}), we may find conditions to see the anomaly originated from the planet in the channel map:
 the planet can approach the emission layer (i.e., $z_{\rm em} \sim \sin\left(\iorb\right)R_0$, where $z_{\rm em}$ is the height of the emission layer).
 Since the minimum mass to see the anomaly is $\mpl/\mstar=3\times 10^{-4}$ when $H_0=0.07$ and $\mpl/\mstar=5\times 10^{-4}$ when $H_0=0.1$, the minimum mass may be given by $\mpl/\mstar \sim 0.5 H_0^3$.

\subsection{Variation of kinematic features in the orbital period} \label{subsec:location_dependence}
\begin{figure}
   \begin{center}
      \resizebox{0.49\textwidth}{!}{\includegraphics{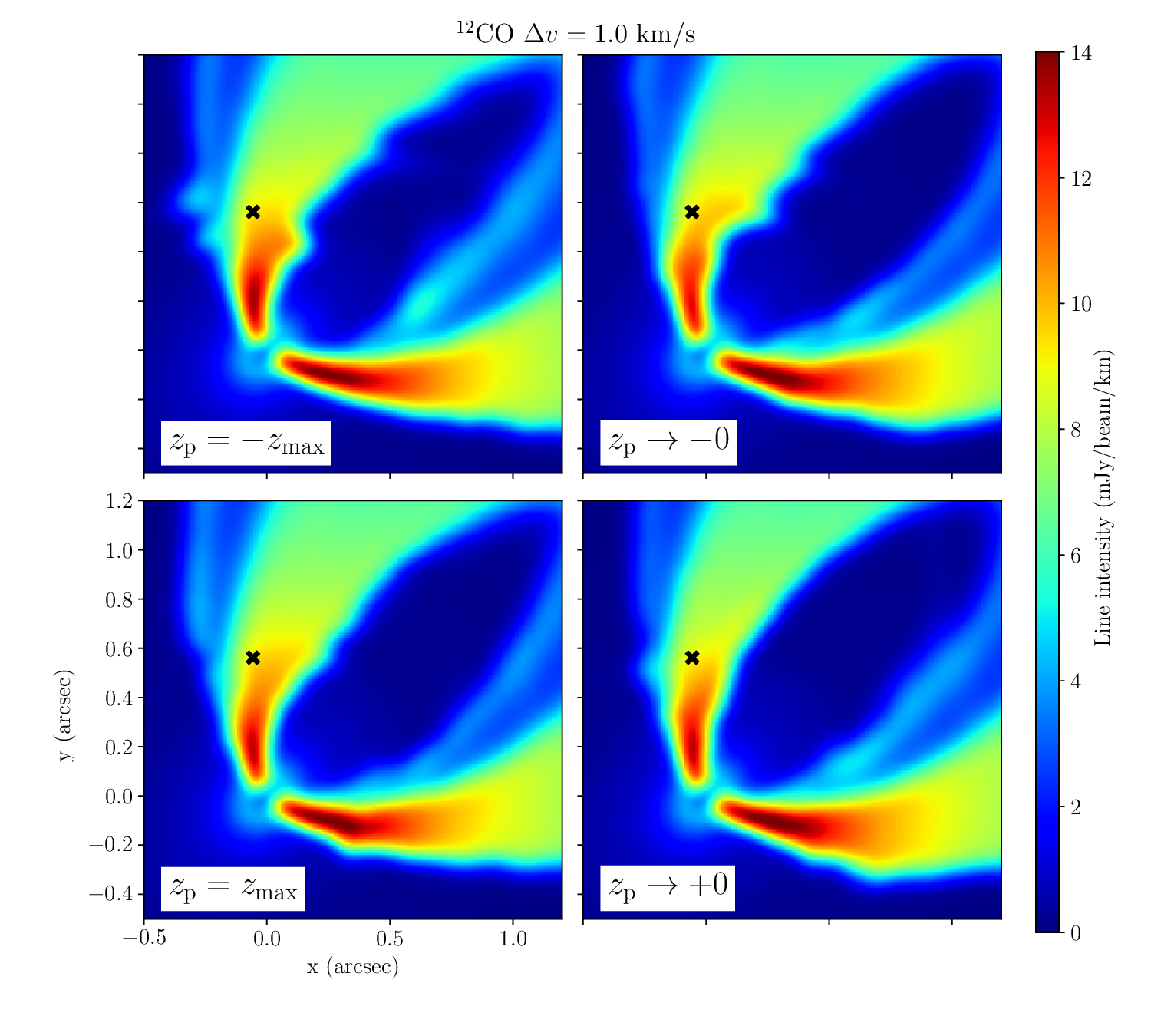}}
   \end{center}
   \caption{
      The variation of the channel map of $\Delta v =1.0$~km/s of \twco~ emission in Run 3 by the meridian location of the planet.
      In the cases labeled by $z_{\rm p}=-z_{\max}$ and $z_{\max}$, the planet is located at the highest altitude of its orbit in $z<0$ or $z>0$ planes (its front/back layer in the case shown in the figure), where $z_{\rm max}/R_0=0.24$ when $\iorb=14^{\circ}$.
      The labels $z_{\rm p} \rightarrow -0$ and $z_{\rm p} \rightarrow +0$ indicate the cases that the planet is located at the midplane ($z_p=0$) from the $z<0$ and $z>0$ layers. 
   \label{fig:co_chmap_varploc}
   }
\end{figure}
\begin{figure}
   \begin{center}
      \resizebox{0.49\textwidth}{!}{\includegraphics{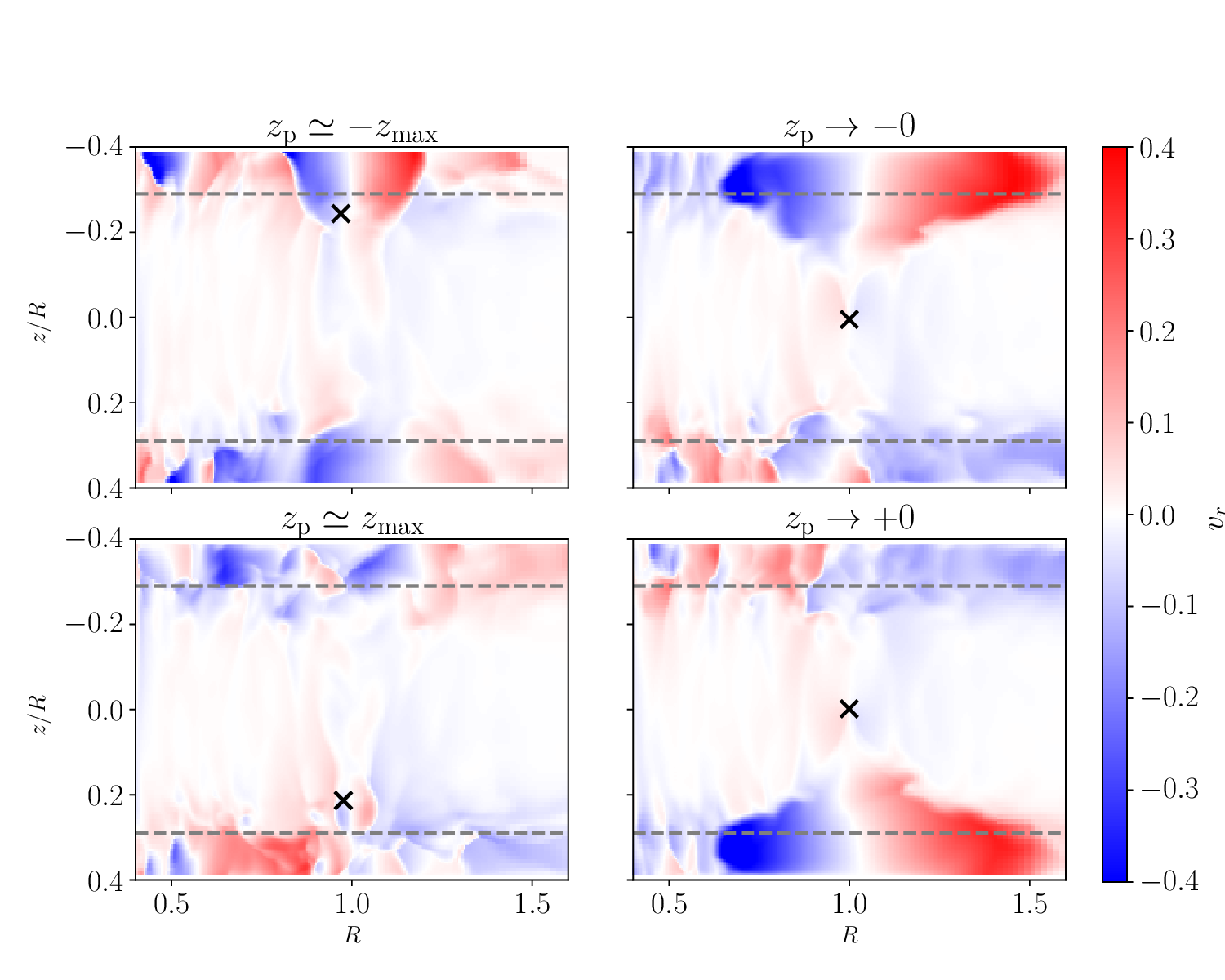}}
   \end{center}
   \caption{
      Distribution of $\vradius$ averaged in the range of $\phi=\phi_{\rm p} \pm R_H/R_0$ in the case shown in Figure~\ref{fig:co_chmap_varploc}.
      The black cross represents the location of the planet, and the horizontal line represents the emission layer of \twco.
   \label{fig:vr_varploc}
   }
\end{figure}
The velocity distribution varies on the meridian location of the planet; hence, the kinetic features in the channel map also change in the orbital period of the planet.
Figures~\ref{fig:co_chmap_varploc} and \ref{fig:vr_varploc} show the channel map of \twco~for $\Delta v =1$~km/s and the distribution of $\vradius$ averaged in the range of $\phi=\phi_{\rm p} \pm R_H/R_0$, respectively.
In the left panels of the figures (the cases labeled by $z_{\rm p}=\pm z_{\max}$), the planet is located at the highest altitude of its orbit in the front ($z<0$)/back ($z>0$ layers). Moreover, in the right panels of the figures, the planet is located at the midplane, but it comes from the front/back layer in the upper/lower panel. 
Significant deviations from the Keplerian rotation are observed when the planet is located in the front layer (in the top panels in Figures~\ref{fig:co_chmap_varploc} and \ref{fig:vr_varploc}). However, the appearances of the deviation are different depending on the location of the planet because the velocity distribution is different (Figure~\ref{fig:vr_varploc}).
Meanwhile, the deviation is much weaker when the planet is located in the back layer because the emission from the back layer that is disturbed by the planet is hidden by the front layer.
\revs{Note that the velocity maps of the left top and left bottom panels in Figure~\ref{fig:vr_varploc} are not symmetric against the midplane, but this asymmetry comes from that the plotting timings are slightly off. The figure that the plotting timing is adjusted is shown in Appendix~\ref{sec:time_corr_fig8}}.

\subsection{Comparison with the case with the coplanar planet} \label{subsec:comp_coplanar_case}
\begin{figure*}
   \begin{center}
      \resizebox{0.98\textwidth}{!}{\includegraphics{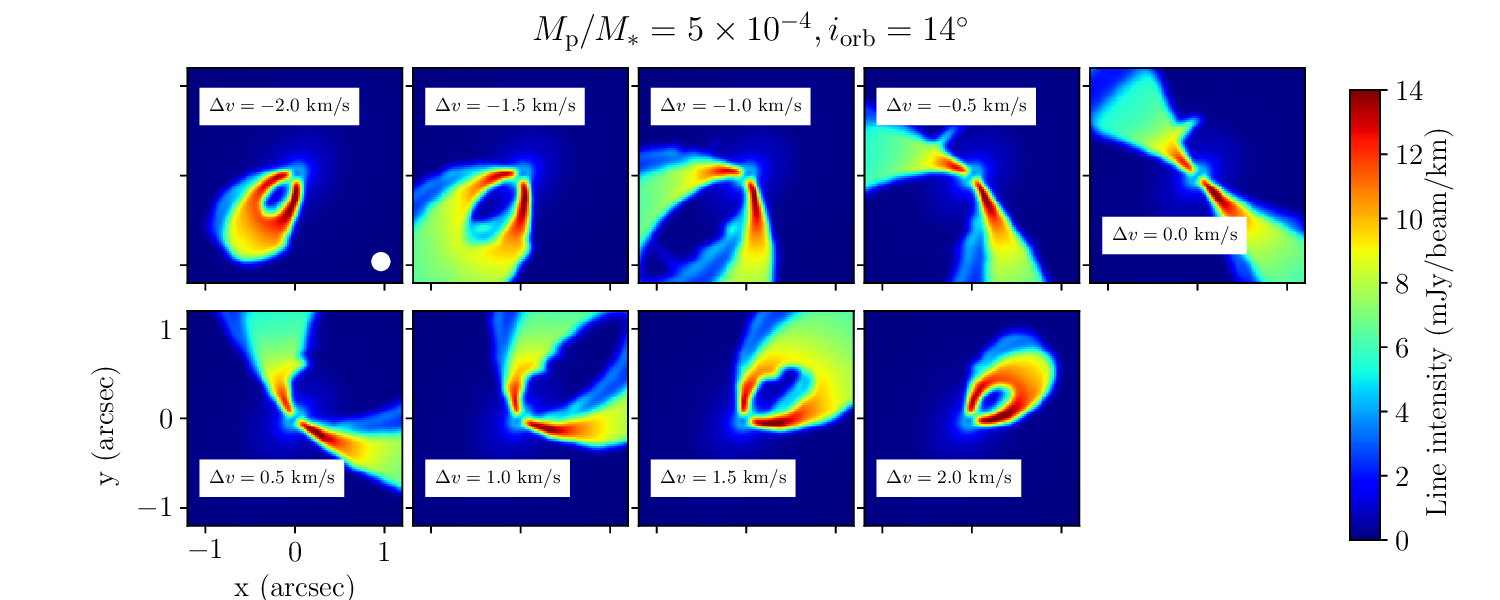}}\\
      \resizebox{0.98\textwidth}{!}{\includegraphics{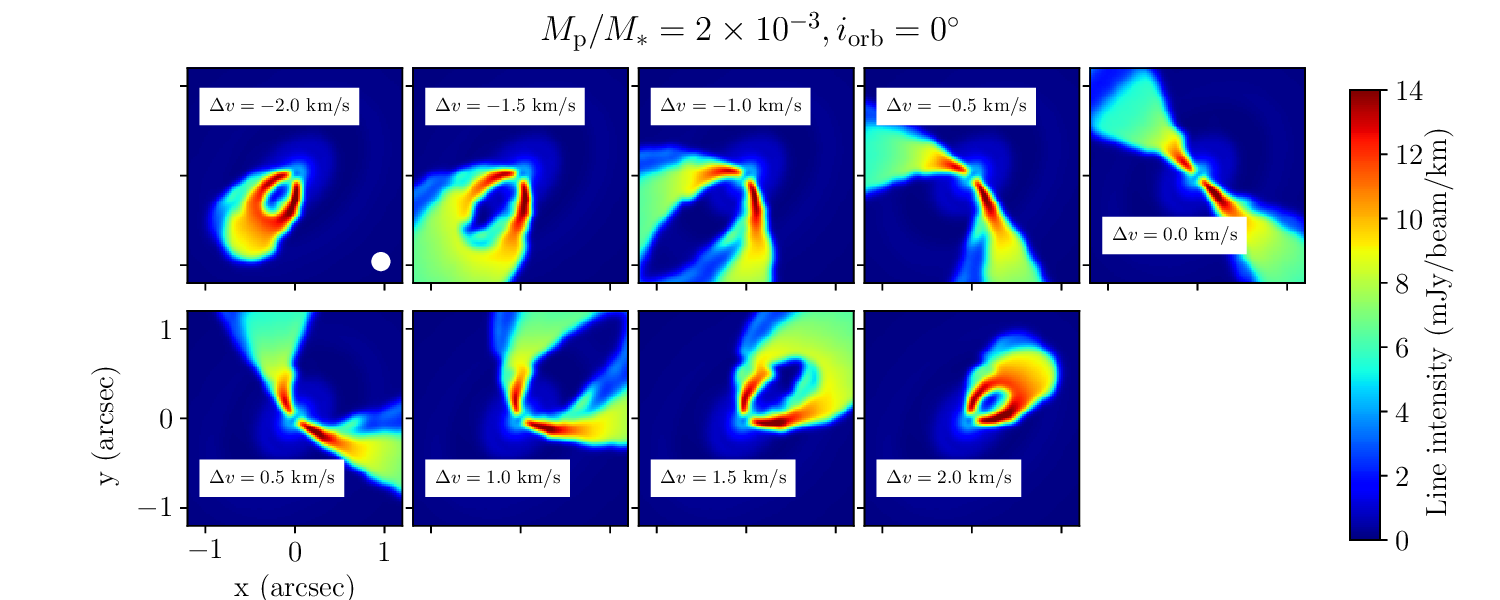}}
   \end{center}
   \caption{
      Series of the channel maps of \twco~J=2--1 emission.
      The upper panels show the channel maps in the case of the inclined planet with $\mpl/\mstar=5\times 10^{-4}$ (Run 3), and the lower panels show these in the case of the coplanar massive planet with $\mpl/\mstar=2\times 10^{-3}$ (Run 11).
   \label{fig:comp_chmap_12co}
   }
 \end{figure*}
 Kinematic planetary features can appear in a series of channel maps, although we focused on that for $\Delta v=1$~km/s in aforementioned subsections.
 In this section, the series of the channel maps for CO isotopologues are shown, comparing the cases with the inclined and coplanar planets. 
 Figure~\ref{fig:comp_chmap_12co} shows a series of channel maps of \twco~emissions in the case of the inclined planet (Run~3) and the massive coplanar planet (Run~11).
 In the case of the massive coplanar planet (lower panels), there are some small deviations from the Keplerian rotation in all the channels compared with those in the case of the inclined planet.
 However, prominent deviations similar with that in the $\Delta v=1$~km/s appear at the similar location in the same channels, such as $\Delta v=0$, 0.5, and 1.5~km/s, in both cases.
 
\begin{figure*}
   \begin{center}
      \resizebox{0.98\textwidth}{!}{\includegraphics{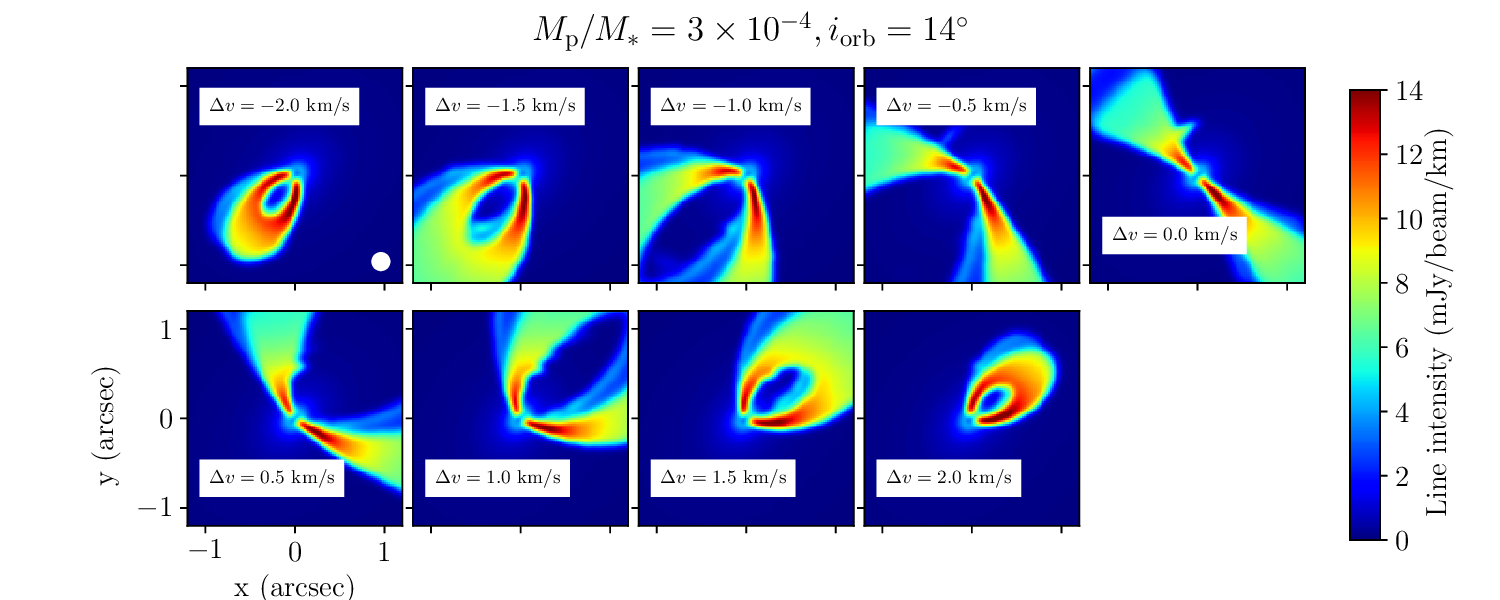}}
   \end{center}
   \caption{
      The same of the upper panels of Figure~\ref{fig:comp_chmap_12co}, but in the case of $\mpl/\mstar=3\times 10^{-4}$ (Run 5).
   \label{fig:chmap_12co_q3e-4}
   }
 \end{figure*}
 As an intriguing feature only appearing in the case of the inclined planet, the tail from the planet location in the $\Delta v=-0.5$~km/s channel can be found.
 However, this tail can be observed only when the planet is located near the midplane.
 Moreover, \rev{the tail is much fainter with the smaller mass of the planet} even if the planet locates around the midplane (Figure~\ref{fig:chmap_12co_q3e-4}). The series of the channel maps in the case of $\mpl/\mstar=3\times 10^{-4}$ (Run 5) is presented in Figure~\ref{fig:chmap_12co_q3e-4}.
 Except for the tail, the structures are similar to these in the case of $\mpl/\mstar=5\times 10^{-4}$, which is also highlighted in Section~\ref{subsec:mass_dependence}.
 Nevertheless, the tail could be evidence of the inclined planet and constrained the mass of the planet, if it is observed.
\rev{However, this tail originates from the gas very close to the planet, with a size comparable to the circumplanetary disk, but our simulation does not resolve the circumplanetary disk (its size is thought to be on the order of $0.1 R_H$). Therefore, this tail structure can be the artifact caused by the unresolved circumplanetary disk.}

\begin{figure*}
   \begin{center}
      \resizebox{0.98\textwidth}{!}{\includegraphics{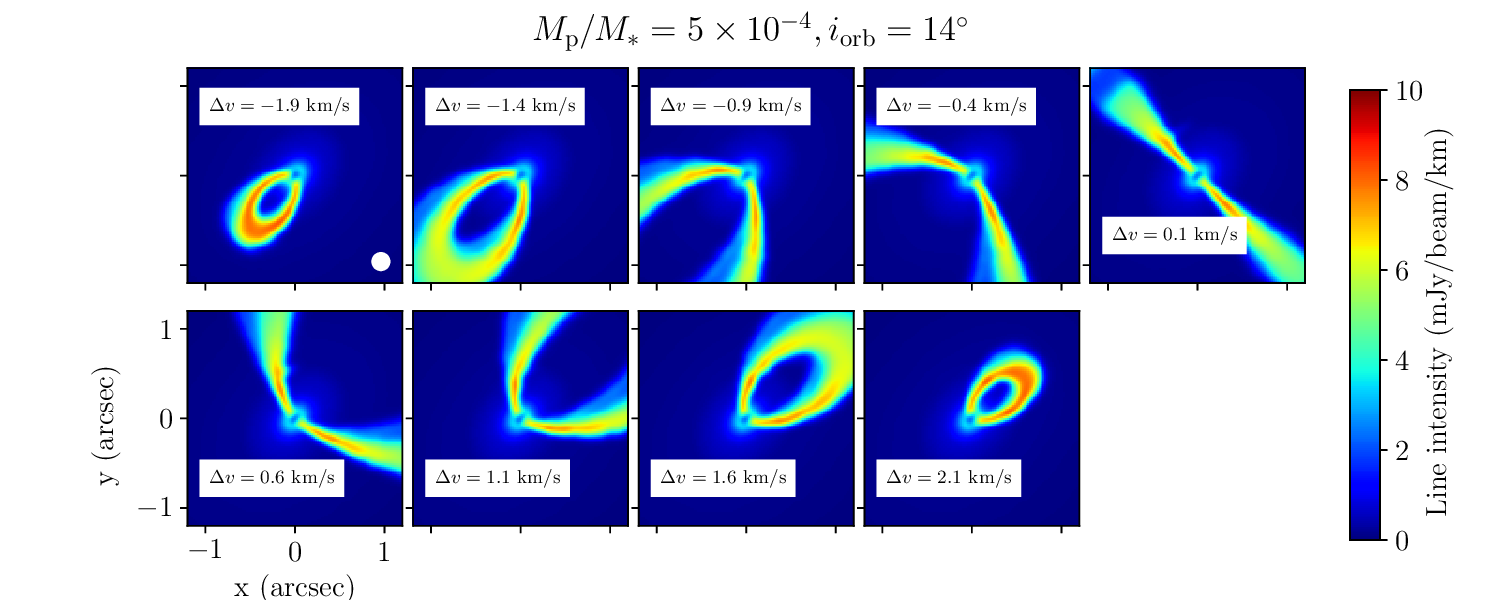}}\\
      \resizebox{0.98\textwidth}{!}{\includegraphics{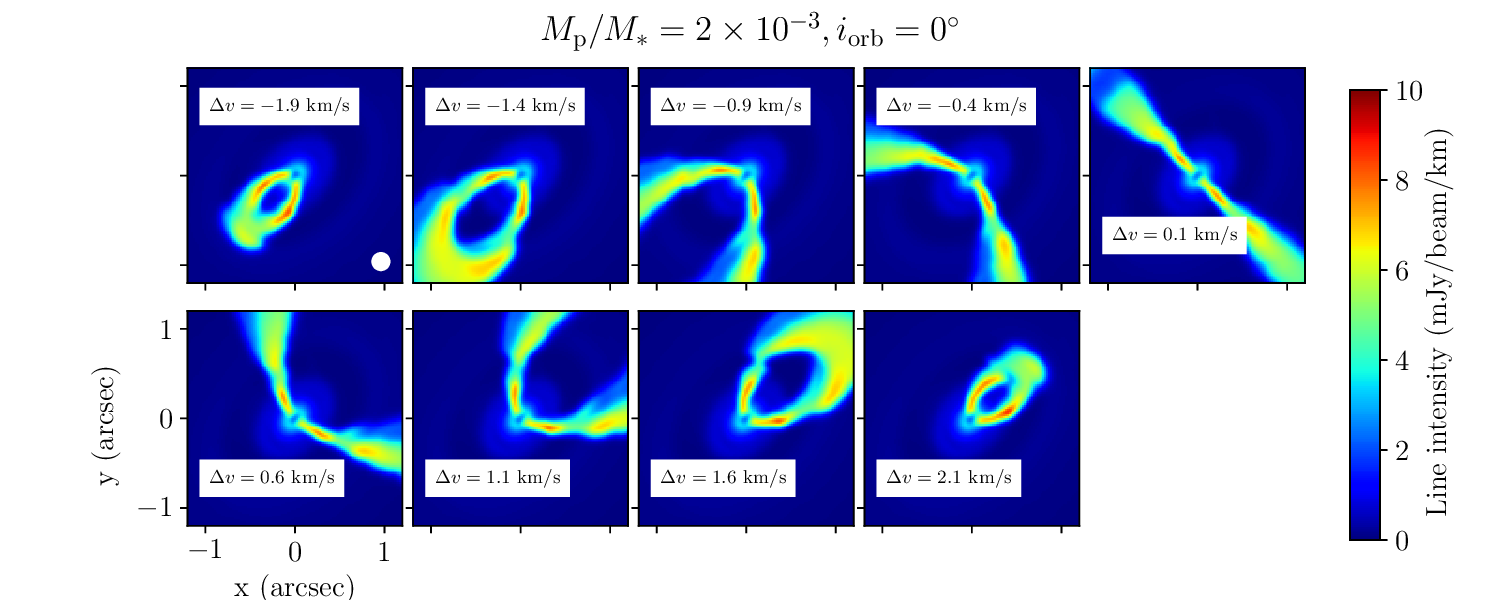}}
   \end{center}
   \caption{
      The same as Figure~\ref{fig:comp_chmap_12co}, but for the \ceio~J=2--1 emissions.
   \label{fig:comp_chmap_c18o}
   }
 \end{figure*}
 Figure~\ref{fig:comp_chmap_c18o} shows the series of channel maps of \ceio~J=2--1 emissions, which is the same cases shown in Figure~\ref{fig:comp_chmap_12co}.
 In the case of the inclined planet shown in the upper panels of the figure, there is almost no deviation from the pattern of the Keplerian rotation, as discussed in Section~\ref{subsec:mass_dependence} and shown in Figure~\ref{fig:cont_coisotops_qvar}.
 In contrast, the prominent gap is observed in all the channels in the case of the massive coplanar planet, as shown in the lower panels of Figure~\ref{fig:comp_chmap_c18o}.
 The difference in the channel map of \ceio~is much distinct as compared with the difference in that of \twco.
 Hence, the observations of \ceio~can be decisive to determine the orbital inclination of the planet.

\begin{figure*}
   \begin{center}
      \resizebox{0.98\textwidth}{!}{\includegraphics{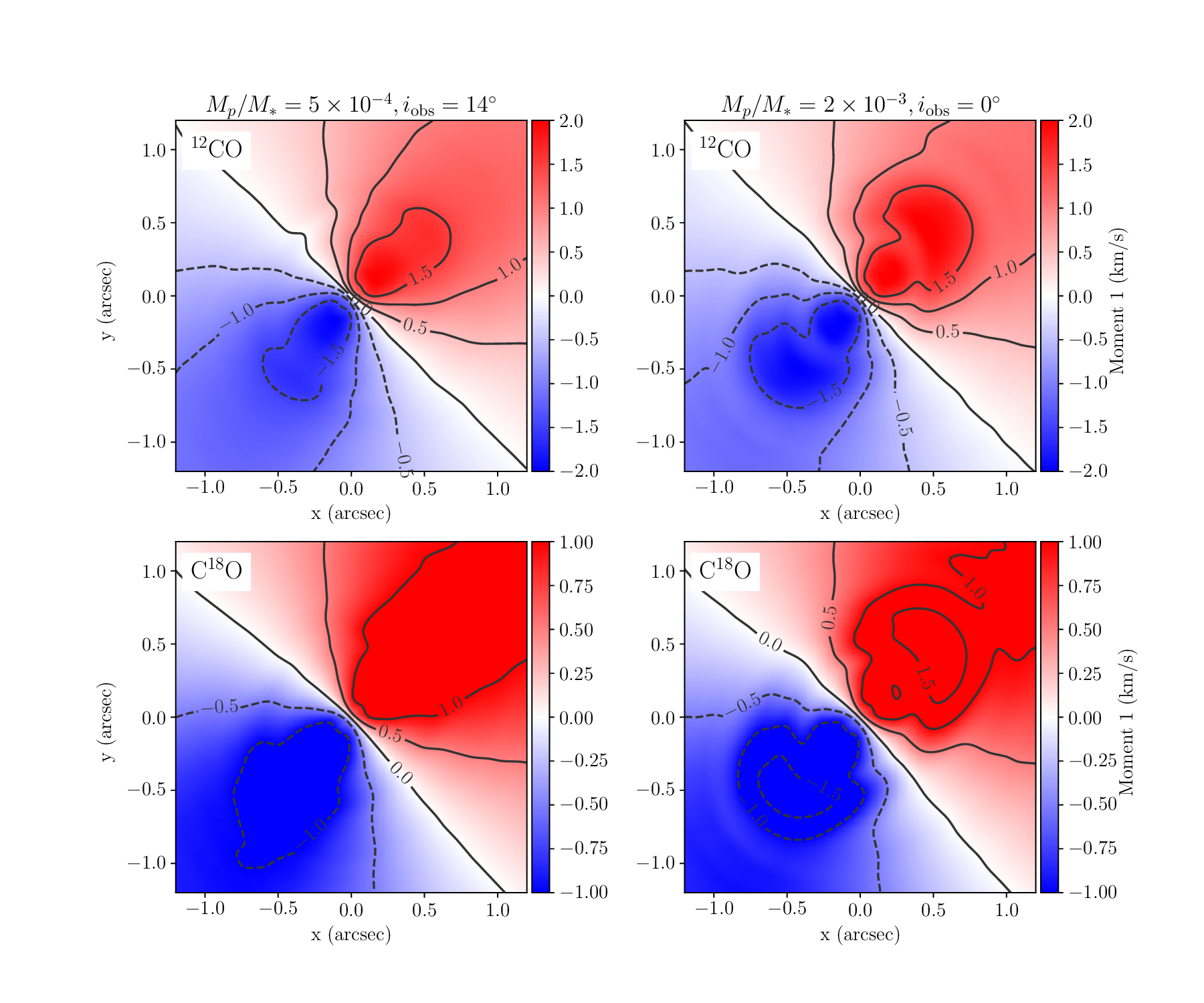}}
   \end{center}
   \caption{
   Moment~1 maps for ${}^{12}$CO (upper panels) and C$^{18}$O (lower panels) with $M_p/M_{\ast}=5\times 10^{-4}$ (Run~3, left panels) and $M_p/M_{\ast}=2 \times 10^{-3}$ (Run~11, right panels).
   \revs{The contours denotes the lines of the moment~1 being equal to $0$~km/s, $\pm 0.5$~km/s, $\pm 1.0$~km/s, and $\pm 1.5$~km/s}.
   }
   \label{fig:mom1}
 \end{figure*}
The \revs{trends} appearing in the channel maps shown above also appear in the moment~1 maps shown in Figure~\ref{fig:mom1}.
As similar to Figure~\ref{fig:comp_chmap_12co}, one finds the deviation from the Keplerian rotation in the $^{12}$CO maps (the upper panels of Figure~\ref{fig:mom1}).
On the other hands, in the C$^{18}$O maps, only a weak deviation can be found in the case of Run~3, while the case of Run~11 shows the deviation as strong as that shown in the $^{12}$CO map.
This is because the inclined planet does not perturb around the midplane and the perturbation is weaker as closer to the midplane, while the massive co-planer planet affects both the midplane and upper layers, as shown above.

\section{Discussion} \label{sec:discussion}
\subsection{Implication to observations} \label{subsec:implications}
When the planet has an inclined orbit, the kinematic features in the channel map has several differences from those induced by the coplanar planet, as presented in Section~\ref{sec:results}.
First, the kinematic features induced by the planet with an inclined orbit is significant in common isotopologues, such as \twco~, because the planet perturbs stronger in a higher altitude (Figure~\ref{fig:velocity_dist_q5e-4}).
For the rarer isotopologues, such as \thco~ and \ceio, only much weaker features are observed (Figure~\ref{fig:cont_coisotops_qvar}).
In contrast, in the case of the coplanar planet, a gap feature can be seen and is clear for rarer isotopologues, because they trace a closer location to the midplane with more decreasing gas density.
The dependence of kinematic features on isotopologues can distinguish the cases with the planet coplanar and inclined orbits.
The gap is narrower and shallower when these are formed by the inclined planet, as compared with those induced by the coplanar planet.

\rev{
 Moreover, we pointed out that, in the outer region of the disk with the disk aspect ratio, the inclined planet can make significant kinematic planetary features in the channel map of \twco~ with no visible gap in dust emission (Figure~\ref{fig:cont_coisotops_h01_ivar}).
 If only the kinematic planetary signature is observed with no dust gap, it may indicate a relatively small planet in the outer region.

 The appearance of kinematic planetary features varies relevant to the vertical location of the planet (Figure~\ref{fig:co_chmap_varploc}).
 Kinematic planetary features can be changed because of the azimuthal rotation of the planet.
 If it is formed by the inclined planet, an additional time variation due to the variation of the vertical location of the planet should be observed.
 Moreover, the planet in the front layer from the observer is likely to be observed because the planet in the back layer produces the weak features in the channel map (Figure~\ref{fig:co_chmap_varploc}).
 That is, half of the planets are hardly detectable.
 This bias should be considered when the planetary population from the observations of the kinematic planetary feature is discussed.
 }

 \rev{We should note that in this paper we have fixed the CO/H$_2$ ratio, the dust/gas ratio, and the H$_2$ gas density itself, and we have also ignored the freezing of CO at very low temperatures, but these quantities determine the height of the emitting layer and thus also affect the appearance of kinetic features.
 In a less massive disk than our model in this paper, for instance, the emission layer of $^{12}$CO lies closer to the midplane, and the neck structure can be seen in $^{12}$CO channel map in this case, and planetary features of CO emission are fainter when the freezing of CO is significant.
 Moreover, the degree of the dust-settling and size distribution of the dust grains can affect the disk temperature (e.g., \cite{Zhang2021}), which also affects the height of the emission layer.
 For direct comparison with the observation and numerical simulations, we should consider non-isothermal equation of state because it changes the shape of the match cone formed by the planet.
 With such the uncertainly, however, comparing the channel maps of different isotopologues, we may distinguish the features induced by the inclined planet and co-planer planet.
 We discuss the case of HD~163296 in detail as an example, below. 
 }

 \cite{Pinte2020} has reported nine disks with significant anomaly from the Keplerian rotation in the channel map of \twco~emission, among the DSHARP sample.
 According to the results of hydrodynamic simulations with the coplanar planet, they suggested that a massive planet with $\gtrsim 1 M_J$ is required to form 
 the observed anomalies from the Keplerian rotation, which is referred as velocity kinks.
 However, they also reported that the planet mass estimated from the gap shape (i.e., width and depth) is $\sim 0.1$ times smaller than that suggested by the velocity kink.
 This discrepancy of the planet masses estimated from the gap and velocity kink may be resolved when the inclined planet is considered, as discussed in Section~\ref{sec:results}.
 If the observations of other isotopologues or other transitions are provided, we could determine which is more suitable to the observation between the coplanar and the inclined planets.

\begin{figure*}
   \begin{center}
      \resizebox{0.98\textwidth}{!}{\includegraphics{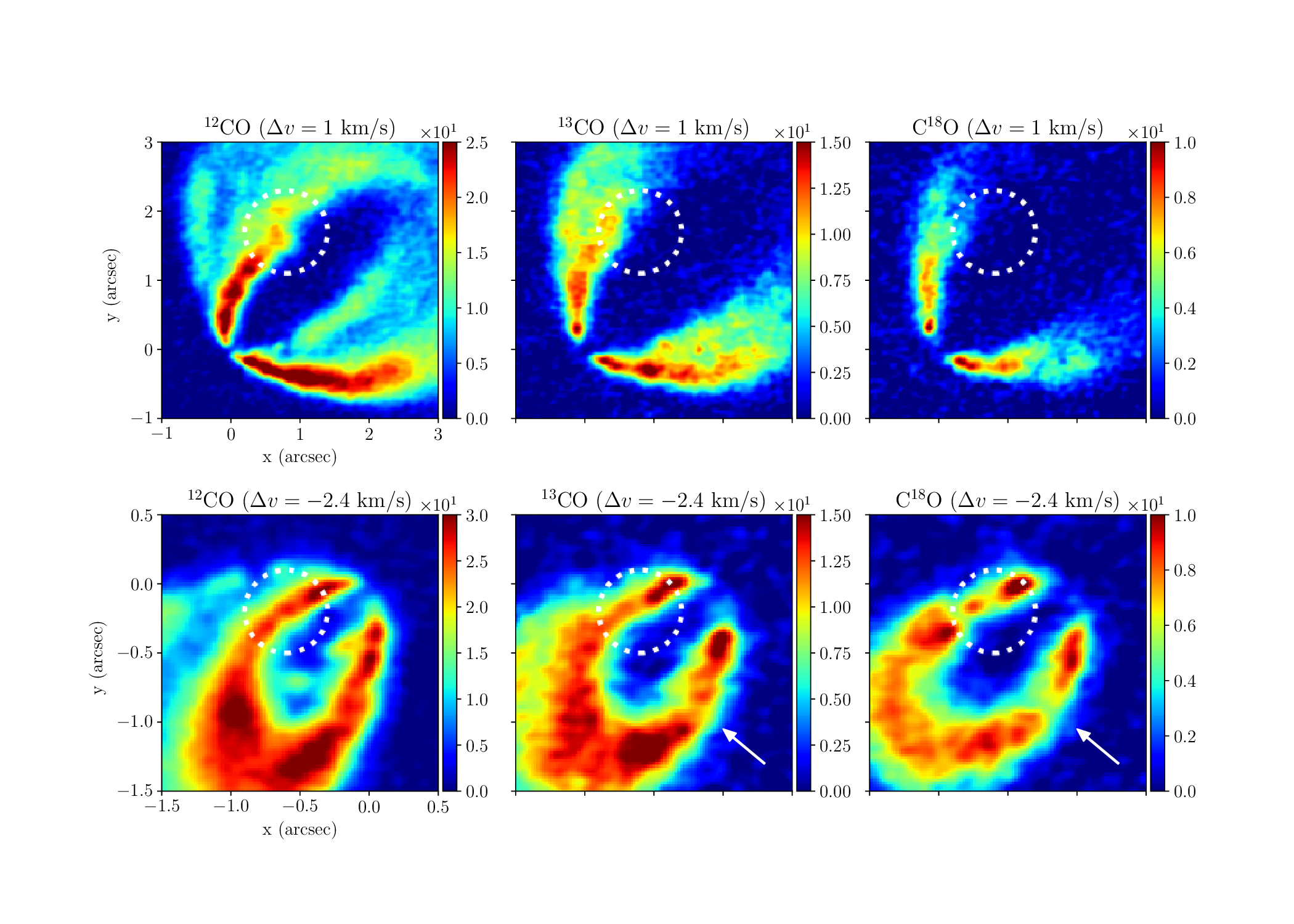}}
   \end{center}
   \caption{
      A part of the channel map of \twco~ (left), \thco~ (middle), and \ceio~ (right) for the disk of HD~163296 is provided by MAPS \citep{MAPS1}.
      The upper and lower panels show the channel map with $\Delta v=1$~km/s and $\Delta v=-2.4$~km/s, respectively.
      The white circles indicate the locations of the velocity kink reported by \cite{Pinte2020}.
      The white arrows in the lower middle and right panels represent the neck and gap structures in the opposite side of the rotation pattern against the planet location.
      The system velocity of HD~163296 is 5.8~km/s \citep{MAPS5}.
   \label{fig:maps}
   }
 \end{figure*}
 Figure~\ref{fig:maps} shows the part of the channel maps for \twco~, \thco, and \ceio, for the disks of HD~163296, which is provided by MAPS \citep{MAPS1}.
 In the disk of HD~163296, \cite{Pinte2020} found two localized velocity kinks: one is seen in the $\Delta v=1$~km/s map at 2.2 arcsec (220~au) (called the outer kink), and the other is seen in the $\Delta v=-2.4$~km/s map at 0.67 arcsec (67~au) (called the inner kink).
 For the outer kink, the significant kink in the \twco~map is found, while there is no significant kink feature in the same location of \thco~kink in the maps of \thco~and \ceio.
 As discussed before, such a dependence on the isotopologues is consistent with that induced by the inclined planet, rather than the coplanar one.
 In contrast, for the inner kink, the neck-like feature and the gap can be observed in the \thco~and \ceio~maps.
 The neck and gap structures can also be observed in the opposite side of the rotation pattern (represented by the arrows in the figure).
 These structures are consistent with that induced by the coplanar planet, as shown in the bottom panels of Figure~\ref{fig:cont_coisotops_qvar}.
 We pointed out that the point-like emission within the white circle in the \ceio~map with $\Delta v=-2.4$~km/s, as shown in the lower right panel of Figure~\ref{fig:maps}).
 The location of the point-like emission may be consistent to the location of the planet estimated by \cite{Pinte2020}.
 \rev{
 If the planet at 220~au has the inclined orbit, the density waves induced by the planet is weaker than that estimated with the co-planer planet, which could explain that \cite{Speedie_Dong2021} explored the density waves in dust continuum but found no clear detection of any spiral structure around 220~au of HD~163296. 
 }

\subsection{Possible mechanisms to maintain the orbital inclination} \label{subsec:inclination_mechanisms}
The orbital inclination is damped due to disk--planet interaction; however, we did not consider this in our paper. 
In the linear regime, the orbital inclination exponentially decays with time and the damping timescale is much shorter than the orbital migration time (\cite{Tanaka_Ward2004}).
Numerically, \cite{Cresswell_Dirksen_Kley_Nelson2007} and \cite{Bitsch_Kley2011} confirmed that the orbital inclination exponentially decays for the small planet with a small inclination, as predicted by the linear theory, except in the large inclination limit as discussed below.
In this case, the inclination of $\sim 10^{\circ}$ can decay within several ten orbits.
For a massive planet with a deep gap, the damping rate can be longer than that estimated by the linear theory \citep{Bitsch_Crida_Libert_Lega2013,Zhu2019}. However, it is not appropriate with the cases that we treated in this paper because only a shallow gap is \rev{opened by} the inclined planet.
Hence, if the observed features are formed by the inclined planet, the mechanism to maintain the inclination is required.

The inclination may be maintained by the balance between damping and excitation, which are due to the disk-planet interaction and \rev{planet-- planet} interaction, respectively, when the planets are captured by the mean-motion resonance (e.g., \cite{Sotiriadis_Libert_Bitsch_Crida2017}).
Almost no visible gap and no kinetic features are observed when the other planet of the pair has a large inclination, as shown in Figure~\ref{fig:cont_co_q5e-4_ivar}.
\cite{Rein2012} investigated the damping timescale of the inclination for a highly inclined planet by the dynamical friction theory. In addition, he showed that the damping because of the disk--planet interaction is inefficient as the inclination increases.
Moreover, the inward migration is slower with a larger inclination.
When the inner planet has a large inclination, the inclination of the outer planet can be excited up to $\sim 10^{\circ}$, and the inward migration can also slow down due to the interaction with the inner planet, which is consistent with the fact that the kinematic features are frequently observed in the outer region of the disk.
It is difficult to directly detect the inner planet with a large inclination because it makes no visible gap and kinematic features.
However, it may be revealed by the inclination and eccentricity of the outer planet that makes the observed kinematic features because the inclination and eccentricity should be rapidly damped without the inner planet.

\section{Summary} \label{sec:summary}
In this paper, we investigated the kinematic features in the channel maps of CO isotopologues induced by the inclined planet through 3D hydrodynamic simulations.
We demonstrated that the appearance of the kinematic features induced by the inclined planet is different from that induced by the coplanar planet at several points as follows:
\begin{enumerate}
\item 
When the planet has an inclination of $10^{\circ}$ -- $20^{\circ}$, the smaller planet ($\mpl/\mstar = 3\times 10^{-4}$) can make the kinematic features as clear as that induced by the coplanar massive planet ($\mpl/\mstar=2\times 10^{-3}$) in the channel map of \twco~(Figure~\ref{fig:cont_coisotops_qvar}).
\item In the case of the inclined planet, the kinematic features is fainter for rarer isotopologues because the velocity perturbation is weaker at the  position closer to the midplane (Figure~\ref{fig:velocity_dist_q5e-4} shows that the emission layer of the rarer isotopologues lies closer to the midplane (Figure~\ref{fig:tausurf1}). In contrast, in the case of the coplanar massive planet (the bottom panels of Figure~\ref{fig:cont_coisotops_qvar}), the neck and gap structures are observed in the channel maps of the rarer isotopologues.
\item The inclined planet can make the significant kinematic planetary feature with no prominent gap in dust emission in the outer region of the disk with the large disk aspect ratio (Figure~\ref{fig:cont_coisotops_h01_ivar}).
\item The appearance of the kinematic planetary feature induced by the inclined planet varies with time, which is relevant to the vertical location of the planet (Figure~\ref{fig:co_chmap_varploc}). Hence, the time variation of the feature that is not related to the azimuthal rotation would be observed if it is induced by the inclined planet. Moreover, only half the planets are detectable, which is important to consider planetary population, since the planet in the back layer from the observer produces much weaker features as compared to the planet in the front layer. 
\end{enumerate}

As discussed in Section~\ref{subsec:implications}, compared to the case of the coplanar planet, the fact that the inclined planet can make a shallower gap and prominent kinematic features may resolve the discrepancy, which is pointed by the previous works (e.g.,\cite{Pinte2020}), in which the planet mass required to explain the observed kink is $4$--$10$ times larger than that estimated from the observed gap shape.
Moreover, the channel maps of \thco~and \ceio~can provide a clue to distinguish between the cases with the coplanar massive and inclined planets.
From the structures of the channel maps of \twco, \thco, and \ceio~in in the case of HD~163296, we concluded that the outer velocity kink at $2.2$~arcsec is consistent with that induced by the inclined planet.
In contrast, the inner velocity kink at  $0.67$~arcsec has the neck and gap structures in the maps of \thco~and \ceio, respectively, which is consistent with the case with the coplanar planet. 

The orbital inclination of the planet significantly affects the gap shape and kinematic features in the channel map.
Hence, in the future, a more sophisticated model including the planet inclination is required to confirm the properties of the embedded planet.

\begin{ack}
   We would like to thank the anonymous referee for his/her careful review and helpful comments. 
   This work was supported by the JSPS KAKENHI (Grant Nos. 17H01103, 18H05441 and 19K14779).
   Numerical computations were carried out on the Cray XC50 at the Center for Computational Astrophysics, National Astronomical Observatory of Japan.
   This paper uses the following ALMA data: ADS/JAO.ALMA\#2018.1.01055.L. ALMA is a partnership of ESO (representing its member states), NSF (U.S.), and NINS (Japan), together with NRC (Canada), NSC and ASIAA (Taiwan), and KASI (Republic of Korea), in cooperation with the Republic of Chile. The joint ALMA observatory is operated by ESO, auI/NRAO, and NAOJ.
\end{ack}

\begin{software}
ATHENA++ (\cite{Athena++}, https://www.athena-astro.app), RADMC3D version 2.0 \citep{Dullemond_Dominik2005} and radmc3dPy python package (https://www.ita.uni-heidelberg.de/~dullemond/software/radmc-3d), Matplotlib (\cite{Matplotlib}, http://matplotlib.org), NumPy (\cite{NumPy}, http://www.numpy.org).
\end{software}

\appendix
\section{Channel maps in the vertically isothermal disk} \label{sec:channel_maps}
\begin{figure*}
   \begin{center}
      \resizebox{0.98\textwidth}{!}{\includegraphics{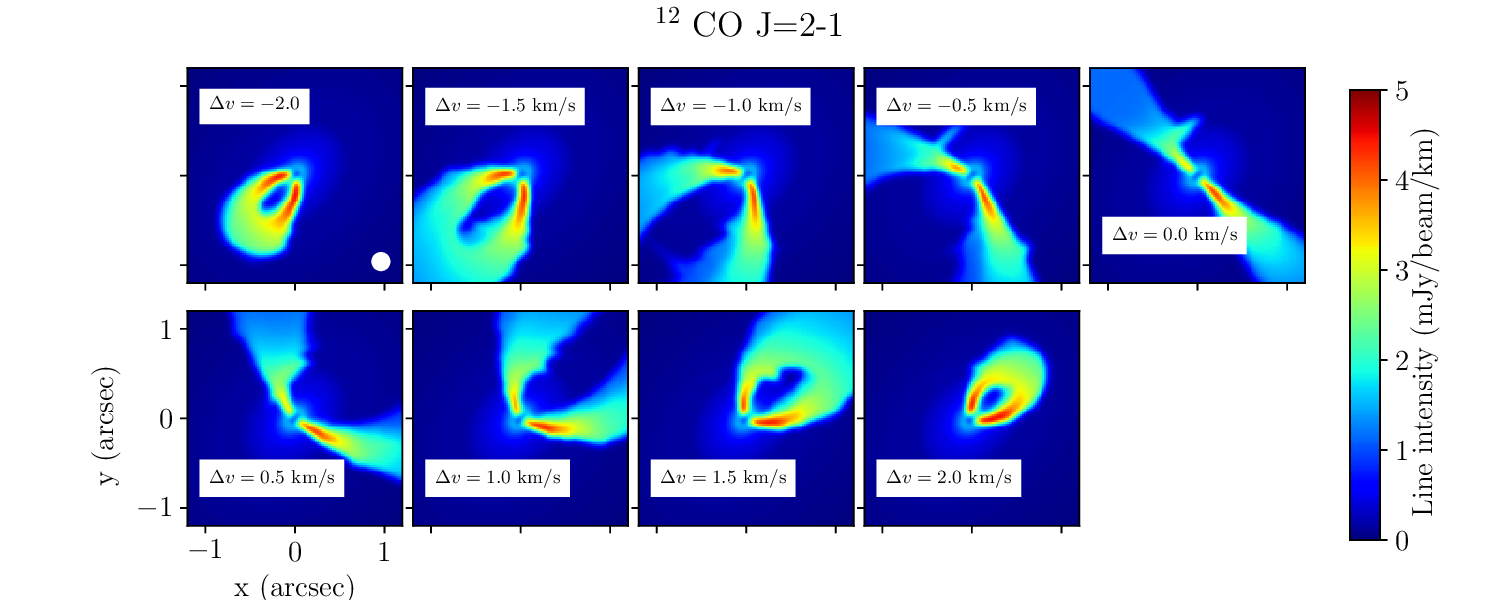}}\\
      \resizebox{0.98\textwidth}{!}{\includegraphics{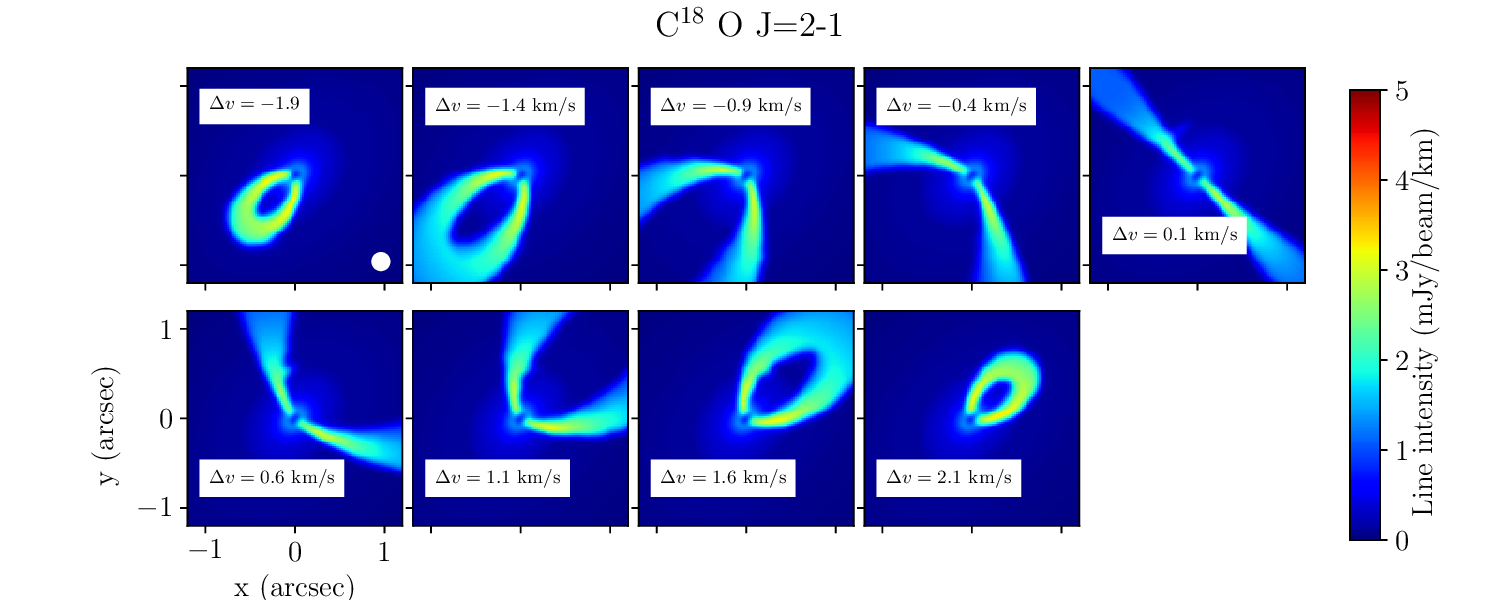}}
   \end{center}
   \caption{
      The series of the channel maps of \twco~(upper panels) and \ceio in the same case are shown in the upper panels of Figure~\ref{fig:comp_chmap_12co}, but the temperature distribution given by Equation~(\ref{eq:csdist}) is adopted.
   \label{fig:chmap_all_q5e-4_viso}
   }
 \end{figure*}
 Figure~\ref{fig:chmap_all_q5e-4_viso} shows the channel map in the same case shown in the upper panel of Figure~\ref{fig:comp_chmap_12co}. However, the temperature calculated from Equation~(\ref{eq:csdist}), vertically isothermal, is adopted.
 Comparing the case that the temperature calculated by the thermal-Monte Carlo calculations (the upper panels of Figures~\ref{fig:comp_chmap_12co} and \ref{fig:comp_chmap_c18o}), the intensities becomes much fainter.
The temperature at an altitude is the same as the midplane temperature in the case shown in Figure~\ref{fig:chmap_all_q5e-4_viso}, while the temperature at an altitude is higher than the midplane temperature when it is given by the thermal Monte Carlo calculation.
 The difference of the intensity is originated from the difference of the temperature around the emission layer.
 However, except for the magnitude of the intensity, the channel maps is considerably similar to each other.
 Although our simulations contain a temperature contradiction adopted in hydrodynamic simulations and radiative transfer simulations, this contradiction is not expected to significantly affect the main results of this paper.

 \section{Disk structure in the case of the massive coplanar planet} \label{sec:coplanar_case}
\begin{figure}
   \begin{center}
      \resizebox{0.49\textwidth}{!}{\includegraphics{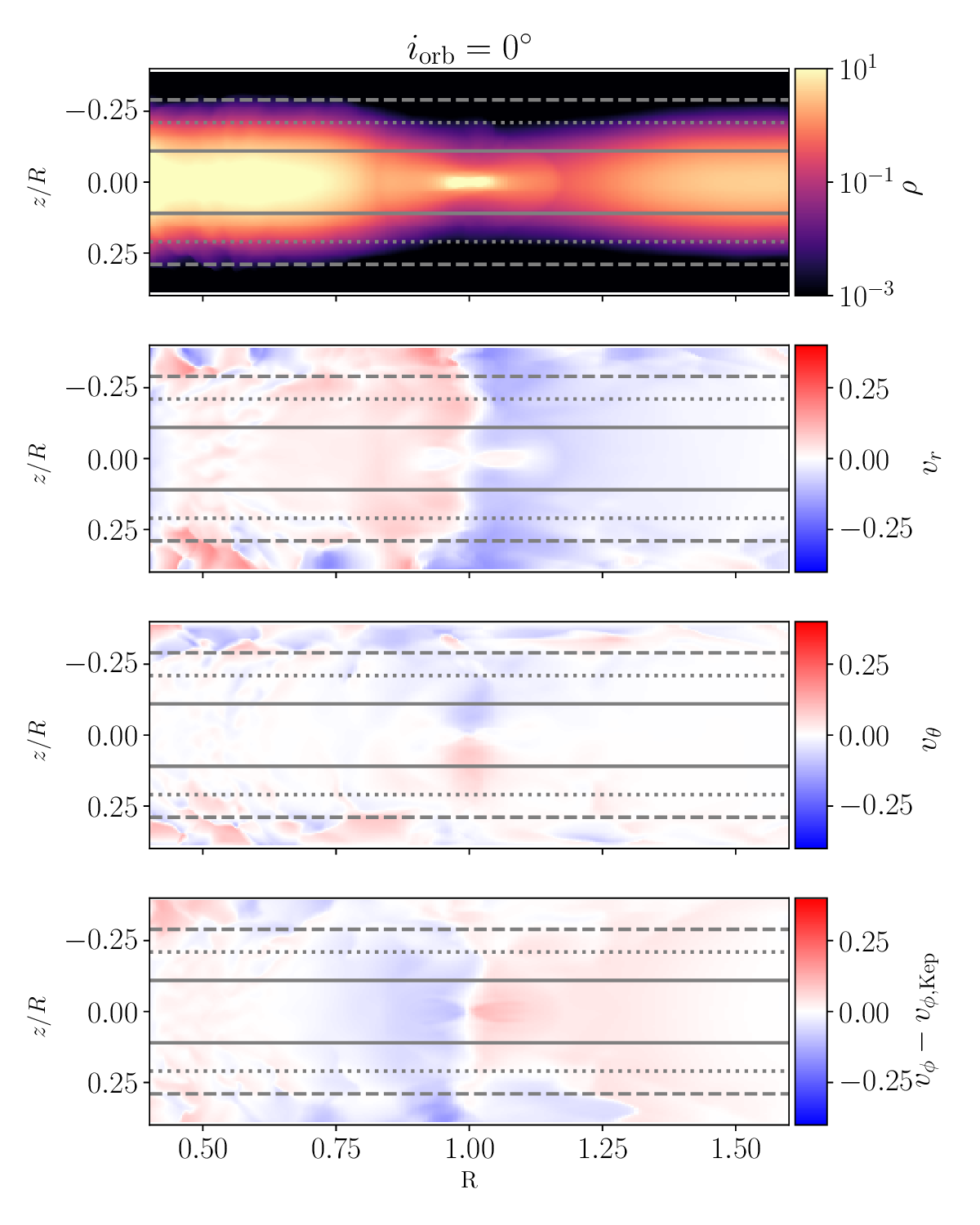}}
   \end{center}
   \caption{
      Distributions of densities and velocities in the case of Run~11.
      The dashed, dotted, and solid horizontal lines represent the emission layers of \twco, \thco, and \ceio~, respectively, in the unperturbed disk (shown in Figure~\ref{fig:tausurf1}).
      The actual heights of the emission layers are affected by gap formation, and these are shown in Figure~\ref{fig:tau1surf_q2e-3}.
   \label{fig:rho_vels_q2e-3}
   }
\end{figure}
\begin{figure}
   \begin{center}
      \resizebox{0.49\textwidth}{!}{\includegraphics{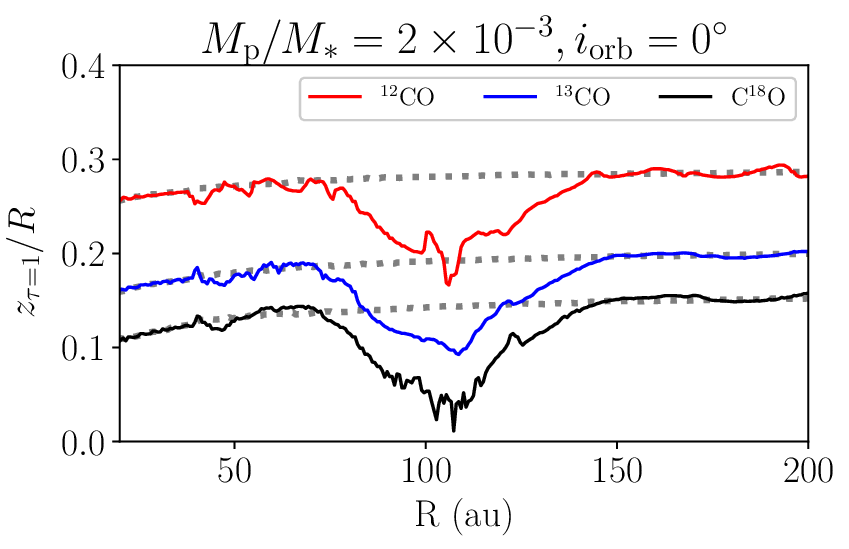}}
   \end{center}
   \caption{
      The same as Figure~ref{fig:tau1surf}, but for the case of Run~11.
   \label{fig:tau1surf_q2e-3}
   }
\end{figure}
 The distributions of densities and velocities in the case of Run 11 is shown in Figure~\ref{fig:rho_vels_q2e-3}, and the heights of the emission layers of CO isotopologues are shown in Figure~\ref{fig:tau1surf_q2e-3}.

\section{Velocity perturbation and the kinematic planetary feature} \label{sec:vel_and_kps}
\begin{figure}
   \begin{center}
      \resizebox{0.49\textwidth}{!}{\includegraphics{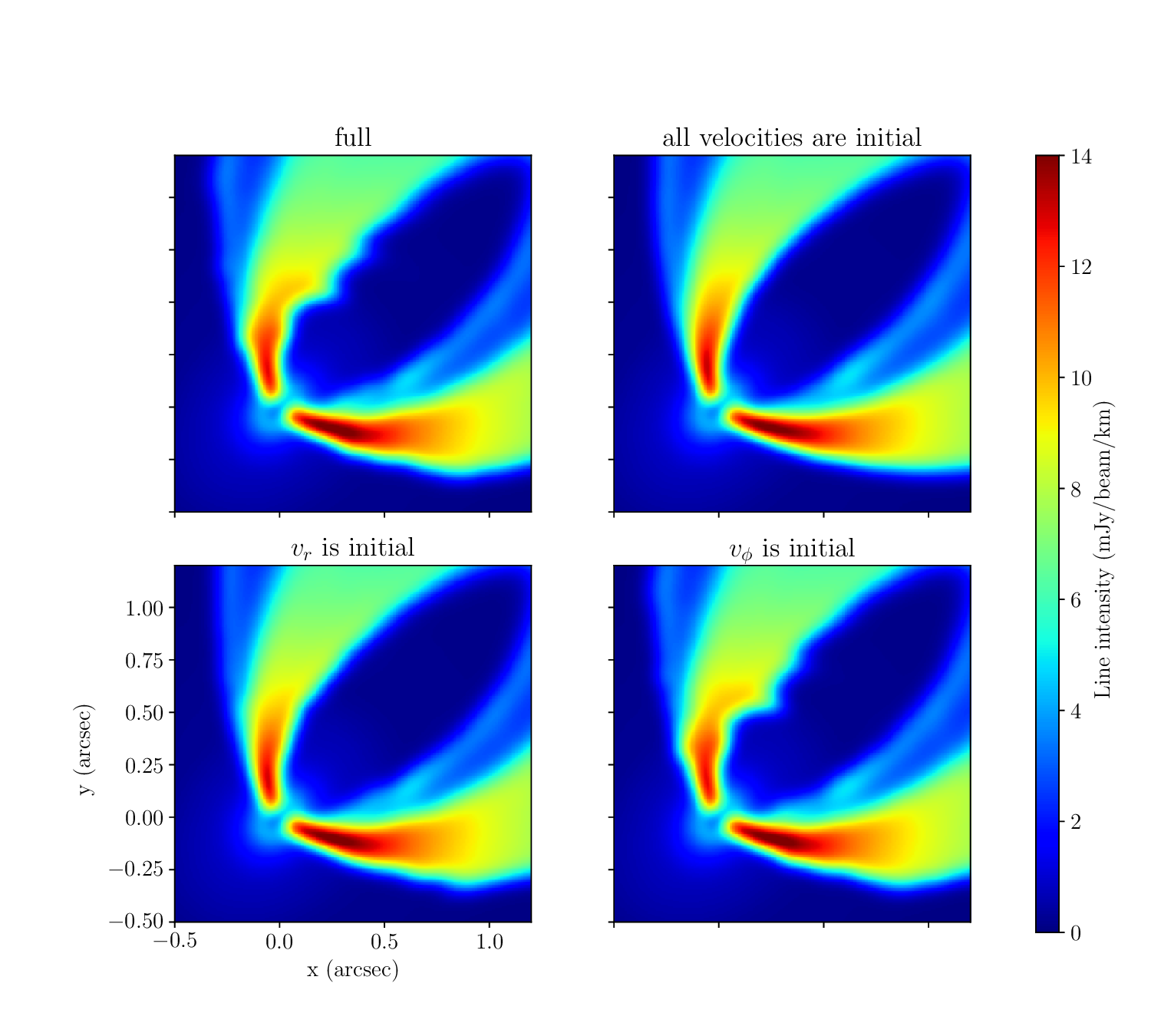}}
   \end{center}
   \caption{
      The channel map of \twco~in $\Delta v=1$~km/s channel in the case of Run~3.
      In the panel labeled as 'full', all physical quantities are fed from the result of hydrodynamic simulation.
      In the panel labeled as 'all velocities are initial' (upper left), all velocities that are given by the initial profile (Keplerian rotation) are adopted instead of those produced by the hydrodynamic simulation, and the lower right and left panels, $\vradius$ and $\vphi$, respectively, are set as the initial profiles. 
   \label{fig:chmap_q5e-4_comp2}
   }
 \end{figure}
In this section, we discuss which components of the velocity mainly contribute to the appearance of the kinematic planetary features.
Considering this, we demonstrated additional radiative transfer simulations in the following three setups: first, the initial profiles of $\vradius$, $\vtheta$, and $\vphi$ were adopted instead of calculating through hydrodynamic simulations. Hence, in this case, only the density was fed from the hydrodynamic simulation.
In the other two setups, we adopted the initial $\vradius$ and $\vphi$, respectively.
The channel maps with these setups are presented in Figure~\ref{fig:chmap_q5e-4_comp2}, when $\mpl/\mstar=5\times 10^{-4}$ and $\iorb=14^{\circ}$ (Run~3).
Based on the figure, when the initial $\vphi$ is applied, the deviation from the Keplerian pattern is observed to be as clear as that in the case where all velocities are fed from the simulation, whereas there is no deviation in the other two cases.
Therefore, the perturbation of $\vradius$ mostly makes the kinematic planetary feature in this case.

\begin{figure}
   \begin{center}
      \resizebox{0.49\textwidth}{!}{\includegraphics{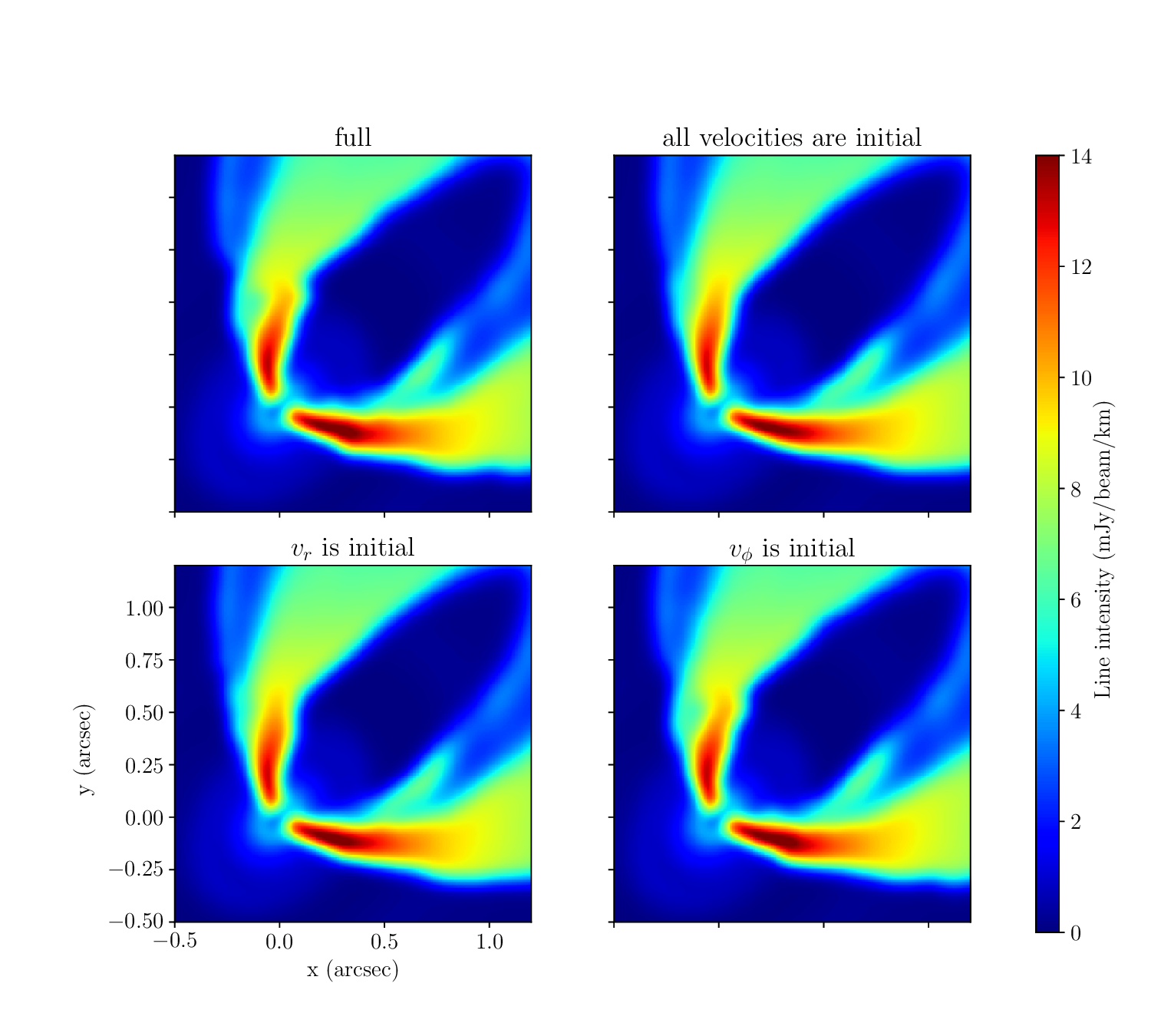}}
   \end{center}
   \caption{
      Similar as Figure~\ref{fig:chmap_q2e-3_comp2}, but for the case of Run~11 (the case of the massive coplanar planet).
   \label{fig:chmap_q2e-3_comp2}
   }
 \end{figure}
 Figure~\ref{fig:chmap_q2e-3_comp2} shows the comparison similar to that presented in Figure~\ref{fig:chmap_q5e-4_comp2} in the case of the massive coplanar planet (Run 11).
 The kink feature around the planet is originated from the perturbation of the $\vradius$, which is similar to Figure~\ref{fig:chmap_q5e-4_comp2}.
 However, even in the case where all velocities are set to be the initial profiles, deviations from the Keplerian pattern can be found, as opposed to the case presented in Figure~\ref{fig:chmap_q5e-4_comp2}.
 Therefore, the perturbation of $\rho$ and $\vphi$ marginally contribute to make the kinematic planetary features and the perturbation of $\vradius$.

\begin{figure}
   \begin{center}
      \resizebox{0.49\textwidth}{!}{\includegraphics{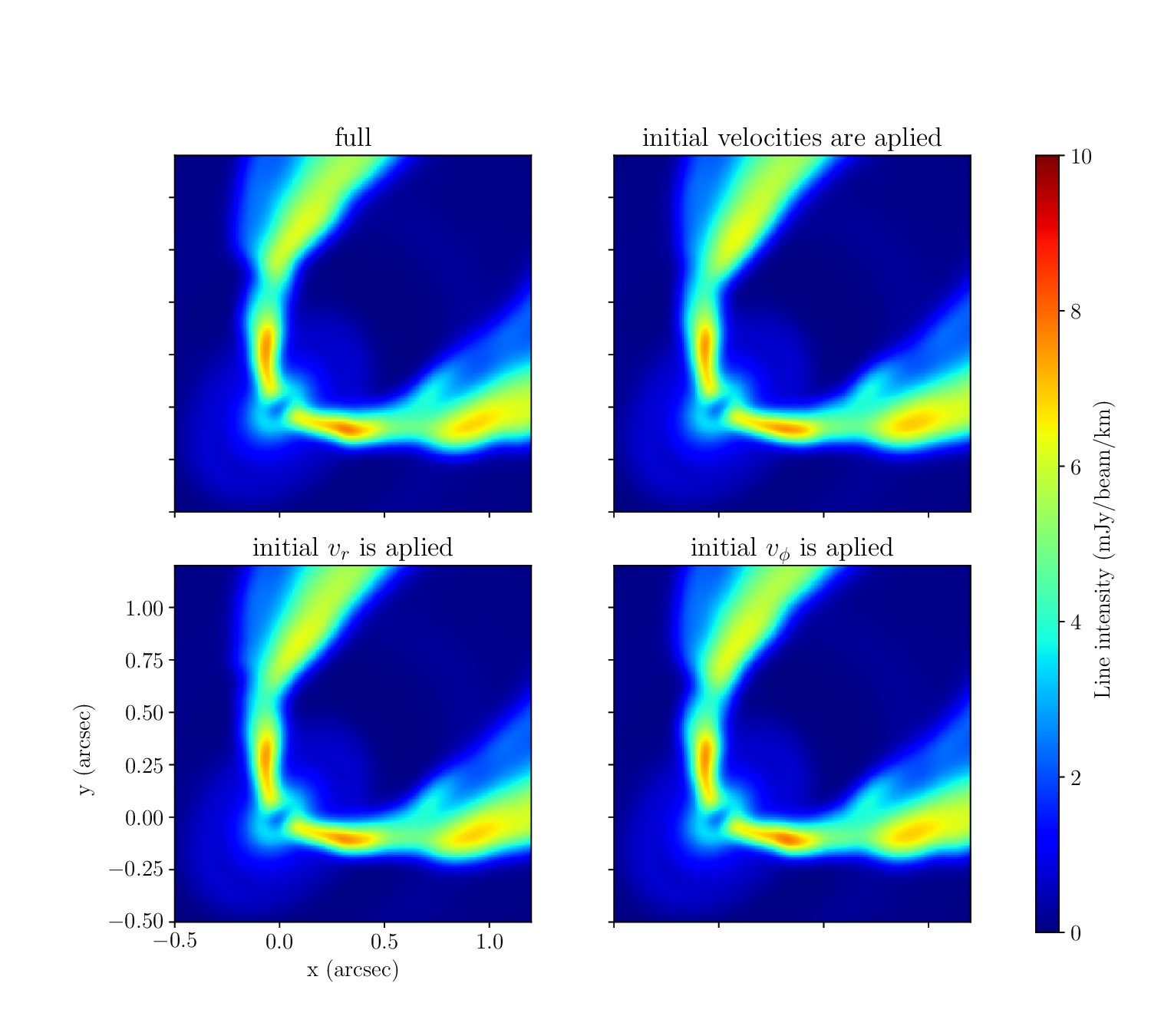}}
   \end{center}
   \caption{
      The same as Figure~\ref{fig:chmap_q2e-3_comp2}, but for \ceio~emissions.
   \label{fig:chmap_q2e-3_comp3}
   }
 \end{figure}
 Figure~\ref{fig:chmap_q2e-3_comp3} is the same as in Figure~\ref{fig:chmap_q2e-3_comp2}, but for \ceio~emission.
 In this case, even the case where all velocities are set as initial profiles, a similar image is produced.
 Hence, in this case, the deviation from the Keplerian pattern is mostly originated from the density perturbation.

\section{Time correction of Figure~\ref{fig:vr_varploc}} \label{sec:time_corr_fig8}
\begin{figure}
   \begin{center}
      \resizebox{0.49\textwidth}{!}{\includegraphics{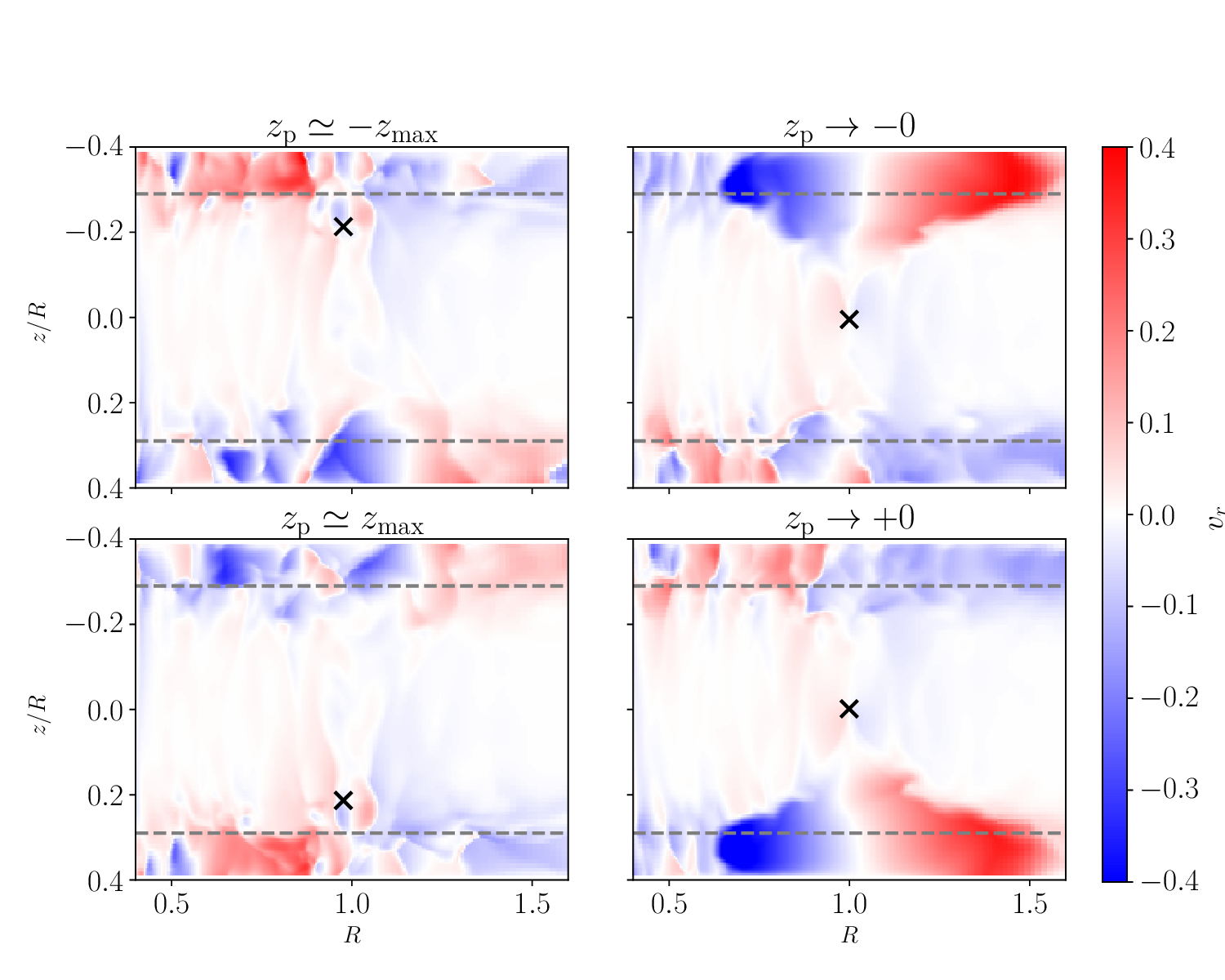}}
   \end{center}
   \caption{
   The same as Figure~\ref{fig:vr_varploc}, but the plotting timing of the left top panel is adjusted to match that of the left bottom panel.
   \label{fig:vr_varploc2}
   }
\end{figure}
\revs{The plotting time of the left top panel in Figure~\ref{fig:vr_varploc} is later in $\pi/\Omegak/6$ that that corresponding to the plotting time of the left bottom panel, and it is fixed in Figure~\ref{fig:vr_varploc2}.}

\end{document}